\documentclass{aa}
\usepackage{natbib}
\usepackage{longtable}

\newcommand{\Msun}{M$_{\odot}$}
\def\elogP{($e, \log P)$}

\def\degr{$^{\circ}$}
\def\Msun{M$_{\odot}$}
\def\SB9{$S\!_{B^9}$}

\bibpunct{(}{)}{;}{a}{}{,}

\def\vec#1{\mathbf{#1}}

\usepackage[dvips]{graphicx}
\usepackage{times}
\newcommand\ignore[1]{} 
\usepackage{amsmath}

\begin{document}
\title{Astrometric orbits of \SB9 stars}
\author{S.~Jancart
\and 
 A.~Jorissen\thanks{Senior Research Associate, F.N.R.S., Belgium}
\and C.~Babusiaux
\and D.~Pourbaix\inst{}\thanks{Research Associate, F.N.R.S., Belgium}}
\institute{
Institut d'Astronomie et d'Astrophysique, Universit\'e Libre de Bruxelles, C.P.~226, Boulevard du Triomphe, B-1050 Bruxelles, Belgium }

\date{Received date; accepted date} 
 
\abstract{
Hipparcos Intermediate
Astrometric Data (IAD) have been used to derive astrometric orbital
elements for spectroscopic binaries from the newly released {\it Ninth
Catalogue of Spectroscopic Binary Orbits} (\SB9). This endeavour is
justified by the fact that (i) the astrometric orbital motion is often
difficult to detect without the prior knowledge of the spectroscopic
orbital elements, and (ii) such knowledge was not available at the
time of the construction of the Hipparcos Catalogue for the
spectroscopic binaries which were recently added to the \SB9\
catalogue.

Among the 1374 binaries from \SB9\ which have an HIP entry (excluding
binaries with visual companions, or DMSA/C in the Double and Multiple
Stars Annex), 282 have detectable orbital astrometric motion (at the
5\% significance level). Among those, only 70 have astrometric orbital
elements that are reliably determined (according to specific
statistical tests), and for the first time for 20 systems. This
represents a 8.5\% increase of the number of astrometric systems with
known orbital elements (The Double and Multiple Systems Annex contains
235 of those DMSA/O systems).

The detection of the astrometric orbital motion when the Hipparcos IAD
are supplemented by the spectroscopic orbital elements is close to
100\% for binaries with only one visible component, provided that the period is in the 50 - 1000 d range and the parallax is $>$ 5
mas.
%in the 50 -- 1000~d period range, when the parallax $\varpi >
%5$~mas, and for binaries with only one visible component (otherwise
%the companion's light would reduce the amplitude of the photocenter
%motion, and make the detection more difficult).
This result is an
interesting testbed to guide the choice of algorithms and statistical
tests to be used in the search for astrometric binaries during the
forthcoming ESA Gaia mission.

Finally, orbital inclinations provided by the present analysis have
been used to derive several astrophysical quantities. For instance, 29
among the 70 systems with reliable astrometric orbital elements
involve main sequence stars for which the companion mass could be
derived. Some interesting conclusions may be drawn from this new set
of stellar masses, like the enigmatic nature of the companion to the
Hyades F dwarf HIP 20935. This system has a mass ratio of 0.98 but
the companion remains elusive.

%One noticeable system is HIP~105860 (= HR~8210), which
%consists of an Am primary, and a hot and very massive white dwarf
%(1.3~\Msun).


}

\maketitle

%
%------------------------------------------------------------------------------
\section{Introduction}
%------------------------------------------------------------------------------
%
The {\it Ninth Catalogue of Spectroscopic Binary Orbits} \citep[\protect\SB9; ]
[ available at {\tt
http://sb9.astro.ulb.ac.be}]{Pourbaix-2004:b} continues the series of
compilations of spectroscopic orbits carried out over the past 35 years
by Batten and collaborators. As of 2004 May 1st, the new Catalogue holds
orbits for 2386 systems. The {\it Hipparcos Intermediate
Astrometric Data} \citep[IAD; ][]{vanLeeuwen-1998:a} offer good prospects
to derive astrometric orbits for those binaries. Astrometric orbits are
often difficult to extract from the IAD without prior knowledge of at
least some among the orbital elements \citep[{\it e.g.,}
][]{Pourbaix-2004:c}. As an illustration of the difficulty, only 45
out of 235 Double and Multiple Systems Annex Orbital solutions
[DMSA/O, see \citet{Hipparcos} and \citet{Lindegren-1997}]
were derived from scratch. For those \SB9\ binaries whose orbit has become
available after the publication of the Hipparcos Catalogue,
new astrometric orbital elements may be
expected from the re-processing of their IAD. This is the major aim of the
present paper, which belongs to a series devoted to the
re-processing of the IAD for binaries \citep{Pourbaix-2000:b,Pourbaix-2003:a}.

One of the major challenges facing astronomers studying binaries and extrasolar planets
is to get the inclination of the companion orbit in order to derive
the component masses. The orbital inclinations will
be provided in this paper for 70 systems (Sect.~\ref{Sect:i}). To get the component masses requires
moreover the system to be spectroscopic binary with 2 observable spectra (SB2). Unfortunately, SB2
systems are not favourable targets to detect their astrometric orbital motion using the IAD.
When the component's brightnesses do not differ much (less than about 1 magnitude), the orbital
motion of the photocenter of the system around its barycenter might not be large enough to allow
detection (see Eq.~\ref{Eq:kappabeta} below). This means that
the astrometric orbit cannot in general be derived from the IAD for SB2 systems (neither can the
component solutions -- DMSA/C -- when available, be reprocessed using the IAD, because the
abscissa residuals of DMSA/C entries turn out to be abnormally large, even for non-binary stars),
thus compromising our ability to derive stellar masses in a fully self-consistent way in the present
paper. This difficulty will be circumvented by the use of the mass -- luminosity relationship for
main sequence stars, thus allowing us to derive at least the {\it companion's} mass 
(Sect.~\ref{Sect:masses}). This
information will then be combined with the position of the system in the eccentricity -- period
diagram to diagnose post-mass-transfer systems (Sect.~\ref{Sect:elogP}).

Another important motivation of the present paper is to test on the IAD, algorithms
designed (i) to detect astrometric binaries and (ii) to determine their orbital parameters in the framework
of the future ESA cornerstone mission Gaia. IAD are indeed very similar to what
will be available at some stage of the Gaia data reduction process.
The fit of an orbital model to the IAD is greatly helped with a partial knowledge of the orbital
elements, coming from the spectroscopic orbit \citep{Pourbaix-2004:c}. In the present
context, orbital elements like eccentricity $e$, orbital period $P$ and one epoch of
periastron passage $T_0$ are provided by the spectroscopic orbits listed in \SB9. With
Gaia, these elements may come (in the most favourable circumstances)
from the spectroscopic orbit derived from the on-board
radial-velocity measurements.

%Moreover, the exploitation of the IAD by the Hipparcos reduction consortia was limited due to
%time constraints in the production of the Hipparcos Catalogue. It may therefore be hoped that
%information -- like astrometric orbits -- is still buried in the IAD and that the
%present work will be able to dig it out. 

 %
%------------------------------------------------------------------------------
\section{The Hipparcos data}
%------------------------------------------------------------------------------
%
During 3 years and for about 118\ts000 stars, the Hipparcos satellite \citep{Hipparcos} measured
tens of abscissae per star, {\it i.e.,} 1-dimensional positions along precessing great circles.
Corrections like chromaticity effects, satellite attitude,
\dots were then applied to these abscissae. It was decided that the residuals $(\Delta v)$ of these
corrected abscissae (with respect to a 5-parameter single-star astrometric model) would be
released together with the Hipparcos Catalogue. They constitute the IAD
\citep{vanLeeuwen-1998:a}. In order to make the interpretation of these residuals unambiguous,
the released values were all derived with the single-star model, no matter what model was used for
that catalogue entry. It is then possible for anybody to fit any model to these IAD to seek
further reduction of the residuals.

\subsection{The orbital model}

The fit of the IAD with an orbital model is achieved through a $\chi^2$ minimization:
\begin{eqnarray}\label{Eq:chi2}
\chi^2 &=& (\Delta v-\sum_k\frac{\partial v}{\partial p_k}\Delta p_k-\sum_i\frac{\partial v}{\partial o_i}o_i)^{\mbox{t}}\nonumber
\\
&&\vec{V}^{-1}(\Delta v-\sum_k\frac{\partial v}{\partial p_k}\Delta p_k-\sum_i\frac{\partial v}{\partial o_i}o_i),
\end{eqnarray}
where $\Delta p_k$ is the correction applied to the original (astrometric) parameter
$p_k$ [where $(p_1, p_2, p_3, p_4, p_5) \equiv (\alpha, \delta, \varpi, \mu_{\alpha^*}, \mu_{\delta})$],
$o_i$ are the orbital parameters and $\vec{V}$ is the covariance matrix of the data. $\Delta v_j,
\partial v_j/\partial p_k,$ and $\vec{V}$ ($j=1,\dots, n; k=1, \dots, 5$) and the Main Hipparcos
solution are provided,
$n$ is the number of IAD available for the considered star [see van Leeuwen \& Evans (1998) for
details].
%The partial derivative of $v$ with respect to the orbital parameter $o$ is given by
%\[
%{\partial v \over\partial o}={\partial v\over\partial p_1}{\partial \xi\over \partial o}+{\partial v\over\partial p_2}{\partial \eta\over \partial o}
%\]
%where $(\xi, \eta)$ are the astrometric coordinates of the photocenter in the plane orthogonal to the line of sight. One simply needs to express $o$ in terms of the orbital parameters and then minimizes Eq.~(\ref{Eq:chi2}).
Equation (\ref{Eq:chi2}) thus reduces to
\begin{eqnarray}
%\label{Eq:chi2}
\chi^2 &=&
(\Delta v-\sum_k\frac{\partial v}{\partial p_k}\Delta p_k-y \frac{\partial v}{\partial p_{1}}-x \frac{\partial v}{\partial
p_{2}})^{\mbox{t}}\nonumber\\
&&\vec{V}^{-1}
(\Delta v-\sum_k\frac{\partial v}{\partial p_k}\Delta p_k-y \frac{\partial v}{\partial p_{1}}-x \frac{\partial v}{\partial p_{2}})
\label{Eq:chi2red}
\end{eqnarray}
where ($x$,~$y$) is the relative position
of the photocenter with respect to the barycenter of the binary system given by
\begin{eqnarray*}
x &=& AX + FY \\
y &=& BX + GY
\end{eqnarray*}
with
\begin{eqnarray*}
X &=& \cos E- e \\
Y &=& \sqrt{1-e^2} \sin E.
\end{eqnarray*}
$A, B, F, G$ are the Thiele-Innes constants (describing the photocenter orbit), $e$ is the
eccentricity and
$E$ the eccentric anomaly.

\subsection{Outliers screening}

Even in the original processing, not all the observations were used to derive the astrometric
solution. Some of the observations were flagged as outliers and simply ignored if their residuals
exceeded three times the nominal ({\it a priori}) error for those measurements. These outlying observations are
identified by lower case 'f' or 'n' flags in the IAD file (instead of upper case `F' or `N' flags,
corresponding to processing by the FAST or NDAC consortium, respectively). Since the model (and
therefore the residuals) is going to be revised, so must be the outliers. Because the
Thiele-Innes model is a linear one (see Eq.~2), its solution is unique and it may therefore
be used to screen out the outliers of the orbital model.

All observations are initially kept. The observation with the largest residual using the orbital
model is removed and the model fitted again without it. If the original residual exceeds
three times the standard deviation of the new residuals, the observation is definitively
discarded (since the number of observations is always less than 300, random fluctuations
should yield less than 1 observation with a residual larger than $3\sigma$). The process is then
repeated with the new largest residual, and so on. Otherwise, the observation is restored and the
whole process is terminated.

A total of 3\,486 observations (out of 84\,766) are thus removed. 60\% of these outliers turn out
to come from the NDAC processing even though the two consortia essentially contribute for the
same amount of data. The percentage of outliers is ten times larger than in the original
Hipparcos processing.

%
%------------------------------------------------------------------------------
\section{The sample}
%------------------------------------------------------------------------------
%

\begin{figure}[htb]
\resizebox{\hsize}{!}{\includegraphics{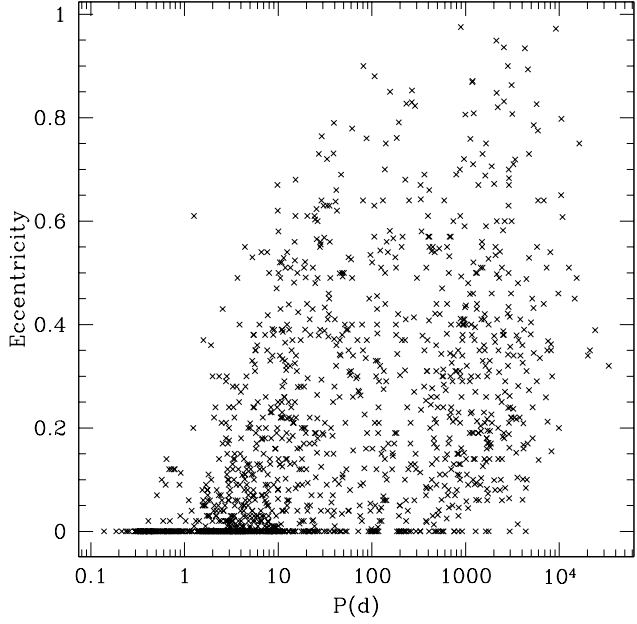}}
\caption[]
{\label{Fig:logP}Period-eccentricity diagram for the selected \SB9 objects with an HIP entry.}
\end{figure}

\begin{figure}[htb]
\resizebox{\hsize}{!}{\includegraphics{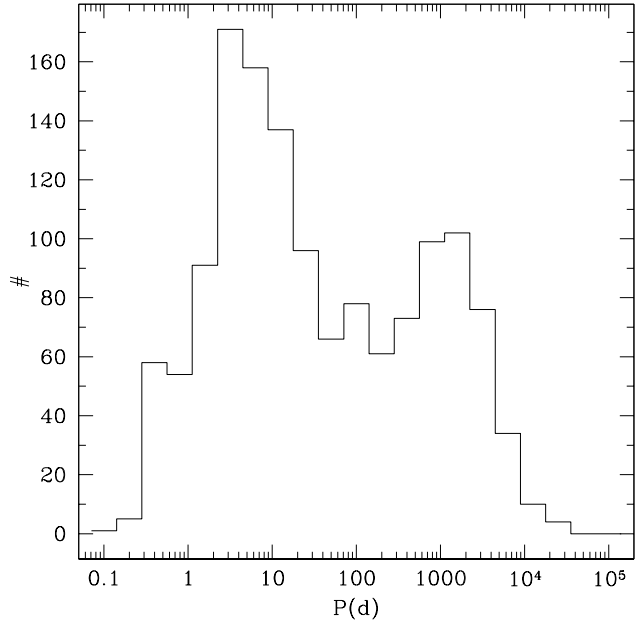}}
\caption[]
{\label{Fig:logP2}Distribution of the orbital periods for the selected \SB9 objects with an HIP entry.}
\end{figure}

Among the $\sim$118\ts000 stars in the Hipparcos catalogue, some 17\ts918 were flagged as double
and multiple systems (DMSA) and 235 of them, the so-called DMSA/O, have an orbital
solution. Our sample consists of the \SB9\ entries with
an HIP number, excluding DMSA/C entries ({\it i.e.,} resolved binaries not suited for IAD
processing). The sample contains 1\ts374 HIP+\SB9 entries which cover an extensive period
and eccentricity range (see Figs.~\ref{Fig:logP} and ~\ref{Fig:logP2}).

%\begin{figure*}[htb]
%\resizebox{0.49\hsize}{!}{\includegraphics{elogp.eps}}
%\hfill
%\resizebox{0.49\hsize}{!}{\includegraphics{period_histo_sb9.eps}}
%\caption[]{\label{Fig:logP}{\bf Left panel: } Period-eccentricity diagram for the selected \SB9 objects with an HIP entry. {\bf Right panel: } Distribution of the period for the selected \SB9 objects with an HIP entry. }%\end{figure*}

Even though a grade characterizes the quality of the spectroscopic orbits listed in \SB9,
those grades were not considered {\it a priori} in the present processing, which uses the most
recent orbit available. The quality of the spectroscopic orbit will be checked at the end of
the process, in the discussion of Sect.~\ref{Sect:wobbledetection} relative to the
detection efficiency of the astrometric wobble.
%
%------------------------------------------------------------------------------
\section{Astrometric wobble detection}
%------------------------------------------------------------------------------
\label{Sect:wobbledetection}
\subsection{Detection assessment}\label{sect:model}

We check whether an orbital motion lies hidden in the IAD using two
mathematically equivalent methods of orbit determination, the Thiele-Innes and Campbell
approaches. In both cases, the eccentricity, orbital period and the time of passage at periastron
are taken from the spectroscopic orbit. For multiple systems, we always use
the shortest period. This choice may not necessarily be the best one, but its validity is anyway
assessed {\it a posteriori} by the `periodogram' test (see below).

In the Thiele-Innes approach, the remaining
four orbital parameters are derived through the Thiele-Innes constants
$A, B, F, G$ obtained from the $\chi^2$ minimization of the linear model expressed by
Eq.~(2). The
semi-major axis of the photocentric orbit ($a_0$), the inclination ($i$), the latitude of the
ascending node ($\Omega$) and the argument of the periastron ($\omega$) (also known as {\it
Campbell's elements}) are then extracted from the Thiele-Innes constants, using standard formulae
\citep{Binnendijk}. In the Campbell approach, on the other hand, two more parameters, $\omega$ and the
semi-amplitude of the primary's radial-velocity curve
$K_1$
are adopted from the spectroscopic orbit. Here, only two parameters of the
photocentric orbit ($i$ and
$\Omega$) are thus derived from the astrometry. This model is non-linear. The Campbell approach implicitly 
assumes that there is no light coming from the companion, since the spectroscopic 
elements constrain a1 according to
\begin{equation}
\label{Eq:a1}
a_1 \sin i =  \varpi  \; \frac{K_1\; P\; \sqrt{1-e^2}} {2 \pi}.
\end{equation}
%and $a_1 = a_0$ if there is no light coming from the companion. 
%This quantity should be compared to the standard deviation of the abscissa residuals.
 The IAD, on the other hand, give access to the {\it photocentric} orbit
characterized by $a_0$, and we assume that a1 = a0. If this assumption does not hold, the solutions derived from the Thiele-Innes and
Campbell approaches will be inconsistent, and will be rejected {\it a posteriori} by the consistency
check described in Sect.~\ref{Sect:orbit}. 

We quantify the likelihood that there is an orbital wobble in the data with a F-test
evaluating the significance of the decrease of the $\chi^2$
resulting from the addition of four supplementary parameters (the four Thiele-Innes constants) in
the orbital model \citep{Pourbaix-2001:b}:
\begin{equation}
\label{Eq:F}
 Pr_2 = Pr(F(4,n-9) > \hat{F}),
\end{equation}
where $ \hat{F} = \frac{n-9}{4} \frac{\chi^2_S - \chi^2_T}{\chi^2_T}$ follows a $F$-distribution with
(4, $n-9$) degrees of freedom, $n$ is the number of
available IAD for the considered star, $\chi^2_T$ and $\chi^2_S$ are the $\chi^2$ values associated
with the orbital and single-star models, respectively.
 $Pr_2$ is the probability that the random variable F(4,n-9) exceeds the given value $\hat{F}$, it is thus 
the first-kind risk associated with the rejection of the null hypothesis `{\em
there is no orbital wobble present in the data}'. The $Pr_2$ test is a $\chi^2$-ratio test; it is therefore insensitive to
scaling errors on the assumed uncertainties. 

An alternative -- albeit non-equivalent -- way to test the presence of an orbital wobble in 
the data is to test whether the four Thiele-Innes constants are significantly different
from 0. The first kind risk associated with the rejection of the null hypothesis `{\em the orbital
semi-major axis is equal to zero}' may be expressed as 
%Getting a substantial reduction of the $\chi^2$ with the Thiele-Innes model does not
%necessarily imply, however, that the four Thiele-Innes constants are significantly different
%from 0. The first kind risk associated with the rejection of the null hypothesis `{\em the orbital
%semi-major axis is equal to zero}' may be expressed as
\begin{equation}
 Pr_3 = Pr(\chi^2_{ABFG} < \chi^2_4),
\end{equation}
where $ \chi^2_{ABFG} = \vec{X}^{\mbox{t}} \vec{C}^{-1} \vec{X}$, $\vec{X}$ is the vector of components
$A,B,F,G$ and $\vec{C}$ is its covariance matrix. $Pr_3$ is thus the probability that $\chi^2_4$, the $\chi^2$ random variable with 
4 degrees of freedom, exceeds the given value $\chi^2_{ABFG}$. The $Pr_3$ test, being based on the  
$\chi^2_{ABFG}$ statistics, is an absolute test, and it is therefore sensitive to possible scaling 
errors on the assumed uncertainties.

\begin{figure}[htb]
\resizebox{\hsize}{!}{\includegraphics{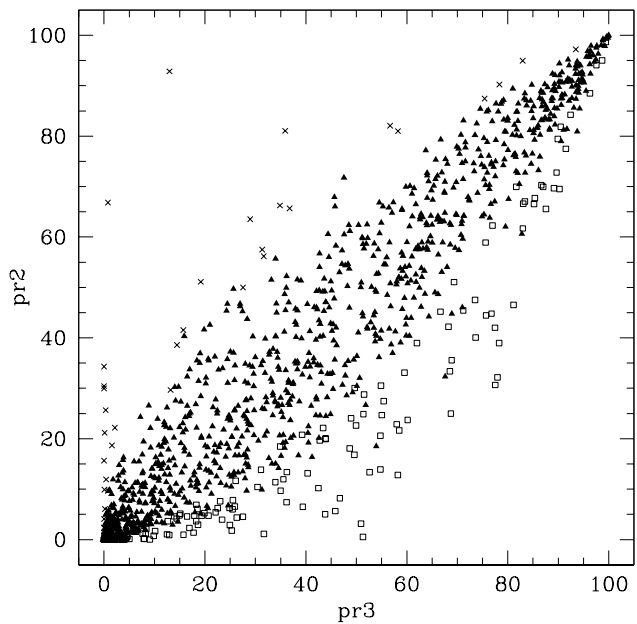}}
\caption[]{\label{Fig:Pr2vsPr3}
Comparison of the $Pr_2$ and $Pr_3$ statistics for the whole sample of 1374 stars, showing that $Pr_2$ and $Pr_3$ are not
equivalent. Crosses correspond
to systems with $F2_{TI} > 2.37$, where $F2_{TI}$ is the goodness-of-fit for the Thiele-Innes model 
(Eq.~\ref{goodness}); open squares correspond to systems with $F2_{TI} < -1.95$.}
\end{figure}

Because $a_0$ vanishes when there is no wobble present in the data (and conversely), it may seem that the $Pr_2$ and $Pr_3$
tests are equivalent (notwithstanding the fact that the former test is relative, whereas the latter is absolute). As revealed by
Fig.~\ref{Fig:Pr2vsPr3}, this is not necessarily so, though, for the reasons we now explain. Since the model is linear, the
equality $\chi^2_{T}=\chi^2_{S}-\chi^2_{ABFG}$ holds. Therefore, $\hat{F} = \frac{n-9}{4} \frac{\chi^2_S - \chi^2_T}{\chi^2_T} =
\frac{n-9}{4} \frac{\chi^2_{ABFG}}{\chi^2_T}$, so that  $Pr_2$ and $Pr_3$ are basically equivalent as long as ${\chi^2_T} \sim
n-9$, {\it i.e.,} when the Thiele-Innes model fits the data adequately. This latter fit may be quantified  
by the goodness-of-fit statistics $F2_{TI}$ \citep{Stuart,Kovalevsky-Seidelmann-04}, defined as:
\begin{equation}\label{goodness}
F2_{TI} = \left(\frac{9 \nu}{2}\right)^{1/2} \left[\left(\frac{\chi^2_T}{\nu}\right)^{1/3} +
\frac{2}{9\nu} - 1 \right],
\end{equation}
where $\nu = n - 9$ is the number of degrees of freedom. If the 
Thiele-Innes model holds, we expect $F2_{TI}$ to be approximately 
normally distributed with zero mean and unity standard deviation.\footnote{The analysis of the single-star fits for the
whole Hipparcos Catalogue reveals that the $F2$ statistics has a mean 0.21 and standard deviation 1.08 \citep{Hipparcos}. This
indicates that the formal errors have been slightly underestimated. Since the same formal errors are used to compute $\chi^2_T$,
the $F2$ statistics for the Thiele-Innes fits has been assumed to have the same parameters as for the 
single-star fits.
Consequently, the 5\% threshold corresponds to $F2 = 2.37$.} Bad fits correspond to large $F2$ values, abnormally good fits to
large negative values. Solutions with $F2 > 2.37$ should
be discarded at the 5\% threshold. 

\begin{figure}[htb]
\resizebox{\hsize}{!}{\includegraphics{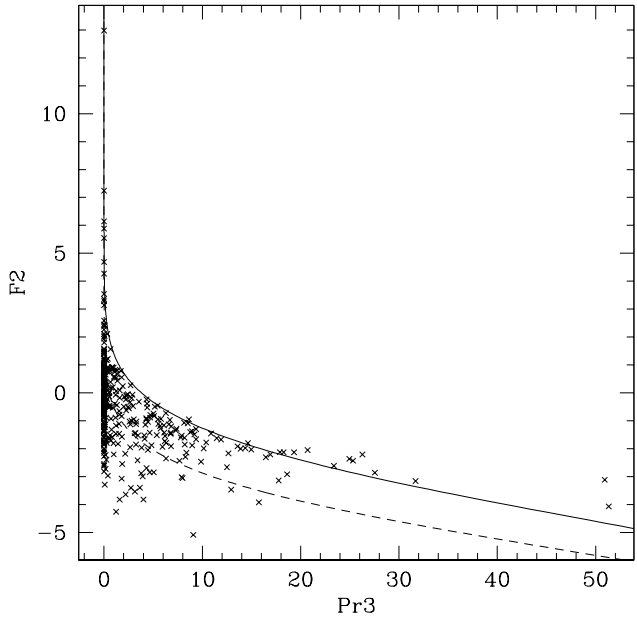}}
\caption[]{\label{Fig:F2vsPr3}$F2$ (goodness of fit) versus $Pr_3$ for systems complying with $Pr_2 < 5$\%. The envelope of these
points is well reproduced with the theoretical curve (solid line) assuming $Pr_2 = 5$\% and 59 observations (which corresponds to
the average number of observations for the considered systems). The dashed line corresponds to $Pr_2 = 1$\%.}
\end{figure}

Fig.~\ref{Fig:F2vsPr3} compares $F2$ with $Pr_3$ and reveals that the two tests are not simple substitutes of one another: there
are systems which fail at the $Pr_3$ test but comply with the $F2$ test and conversely. The situation becomes clearer when one
realizes that the upper envelope corresponds to the condition $Pr_2 < 0.05$, which may be translated into a lower bound on
$\chi^2_{ABFG}/\chi^2_T$: solutions retained by the $Pr_2$ test have large  $\chi^2_{ABFG}/\chi^2_T$ ratios. There are two ways to
fulfill such a condition: 
If $\chi^2_T$ is small ({\it i.e.,} $F2$ is small, or abnormally good fits), then even small $\chi^2_{ABFG}$ values ({\it
i.e.,} large $Pr_3$) comply with the $Pr_2$ test. This explains why the $Pr_2$ test does not eliminate systems with large
$Pr_3$ when their Thiele-Innes fit is abnormally good.  Conversely, if $\chi^2_{ABFG}$ is large ({\it i.e.,} $Pr_3$ is
small), then even large $\chi^2_T$ values ({\it i.e.,} large $F2$ or bad Thiele-Innes fits) comply with the $Pr_2$ test. This
explains why at small $Pr_3$ values, even bad Thiele-Innes fits (large $F2$ values) are retained. This would typically be the case
of a DMSA/X system where the Thiele-Innes model brings a substantial improvement with respect to the single-star model ({\it
i.e.,} $\chi^2_{ABFG}=\chi^2_{S}-\chi^2_{T}$ is large, or $Pr_3$ is small), but the overall quality of the Thiele-Innes fit
remains poor (large $F2$).

In the Campbell approach, the situation is somewhat more complicated since 
the model expressed by Eq.~\ref{Eq:chi2} does not depend linearly upon 
the model parameters $i$ and $\Omega$. Therefore, the quantity $\chi^2_C$ extracted from the minimization of
Eq.~\ref{Eq:chi2} does {\it not} follow a $\chi^2$ distribution with $n-2$ degrees of freedom 
\citep{Lupton-93}. Since the non-linear model may be linearized at the expenses of adding more parameters
(e.g., the coefficients of a Fourier or Taylor expansion), $n-2$ {\it overestimates} the number of 
degrees of freedom \citep{Pourbaix-2005}. Overestimating the number of degrees of freedom affects
all the statistical tests using  the $\chi^2_C$ value.  In particular, the first kind risk $Pr_1$
extracted from an equation similar to Eq.~\ref{Eq:F} (substituting $\chi^2_T$ by $\chi^2_C$)  
is underestimated \citep{Pourbaix-2005}. Since this threshold is used to {\it reject} solutions
which have $Pr_1$ larger than the adopted threshold,
it may nevertheless be used, keeping in mind that {\it not enough} solutions are in fact discarded by the 
$Pr_1$ test. It is very likely, though, that these unacceptable solutions will be screened out by the other tests.
%The same is done for the Campbell approach, yielding the first-kind risk $Pr_1$ \citep{Pourbaix-2001:b}.

The combination of these four statistical indicators allows us to flag 282 stars as astrometric
binaries at the 5\% level ({\it i.e.,} $Pr_1, Pr_2, Pr_3 < 0.05$ and $F2_{\mbox{TI}}< 2.37$) among the
1374 HIP+\SB9\ sample stars defined in Sect.~3.

\subsection{Detection rate}\label{sect:res}

The 282 astrometric binary stars passing the four tests described in
Sect.~\ref{sect:model} at the 5\% level are listed in Table~\ref{Tab:Pr123}. Italicized entries correspond
to the 122 stars passing the $Pr_1, Pr_2$ and  $Pr_3$ tests at the more stringent 0.006\% level 
and $F2 < 2.37$.
These stars thus represent
prime targets for future astrometric observations or, if both components are visible,
interferometric observations
\citep[see also Table~1A of ][]{Taylor-2003}, as they are astrometric binaries, but with orbital
elements not always reliably determined (see Sect.~5).

\begin{table*}[bth]
\caption[]{\label{Tab:Pr123}
The 282 stars flagged as astrometric binaries ($Pr_1,
Pr_2, Pr_3 < 0.05$ and $F2_{\mbox{TI}}< 2.37$; see text). Italicized entries identify
the 122 stars passing the $Pr_1, Pr_2$ and  $Pr_3$ tests at the more stringent 0.006\% level, and $F2_{\mbox{TI}} < 2.37$.}
\begin{tabular}{ccccccccccccc}
\hline
\noalign{\smallskip}
HIP & HIP & HIP & HIP & HIP & HIP & HIP & HIP & HIP & HIP & HIP & HIP & HIP \\
\hline
\hline
\noalign{\smallskip}
443 & 7564 & {\it 13055} & 22961 & {\it 31681} & 43346 & {\it 52419} & 60292 & {\it 73440} & 81170 & 92112 & 99089 & {\it 109176} \\
{\it 677} & 7719 & 14273 & 23402 & 32713 & 43413 & 52444 & {\it 60998} & {\it 74087} & 82706 & 92175 & {\it 99675} & 109554 \\
{\it 1349} & 8645 & {\it 15394} & {\it 23453} & 32761 & 43557 & 52650 & {\it 61724} & {\it 75379} & {\it 82860} & 92177 & {\it 99848} & {\it 110130} \\
{\it 1955} & {\it 8903} & 15900 & 23743 & {\it 32768} & {\it 43903} & 52958 & 62437 & {\it 75695} & {\it 83575} & {\it 92512} & {\it 9996}5 & 110273 \\
{\it 2081} & {\it 8922} & 16369 & 23922 & 33420 & 44124 & {\it 53238} & {\it 62915} & {\it 75718} & 83947 & 92614 & 100384 & 110514 \\
2865 & 9110 & 17136 & 23983 & {\it 34164} & 44455 & {\it 53240} & {\it 63406} & 76267 & 84677 & {\it 92872} & {\it 100437} & 111104 \\
3300 & 9121 & 17296 & {\it 24419} & {\it 34608} & {\it 44946} & 53425 & 63592 & 76574 & 84886 & 92818 & 100738 & {\it 111170} \\
3362 & 10340 & {\it 17440} & 24984 & 35412 & {\it 45075} & 53763 & 64422 & 76600 & {\it 84949} & 92872 & {\it 101093} & 111191 \\
{\it 3504} & {\it 10514} & 17683 & 25048 & 36042 & 45333 & 54632 & {\it 65417} & 77409 & 85829 & {\it 93244} & 101780 & {\it 112158} \\
3572 & 10644 & 17932 & 25776 & {\it 36377} & 46005 & {\it 55016} & 65522 & 77634 & 85985 & {\it 94371} & {\it 101847} & {\it 113718} \\
3951 & {\it 10723} & 18782 & {\it 26001} & 36429 & 46613 & 55022 & 67195 & 77678 & {\it 86400} & {\it 95028} & 102388 & {\it 113860} \\
4166 & {\it 11231} & 19248 & 26291 & 37908 & {\it 46893} & {\it 56731} & {\it 67234} & 77801 & {\it 86722} & {\it 95066} & {\it 103519} & {\it 114313} \\
4252 & {\it 11349} & {\it 20070} & {\it 27246} & {\it 38414} & 47205 & {\it 57791} & {\it 67480} & 78689 & {\it 87895} & 95176 & 103722 & {\it 114421} \\
4584 & 11380 & {\it 20087} & 28537 & 39341 & {\it 47461} & 58590 & {\it 67927 }& {\it 79101} & {\it 88788} & {\it 95575} & {\it 103987} & {\it 116478} \\
4754 & 11465 & 20284 & 29276 & {\it 39424} & 49561 & 59148 & {\it 68072} & 79195 & 88946 & 95820 & 105017 & {\it 116727} \\
4843 & 11843 & 20482 & 29740 & {\it 39893} & {\it 49841} & 59459 & {\it 68682} & {\it 79358} & 89773 & 95823 & 105432 & {\it 117229} \\
{\it 5336} & {\it 11923} & {\it 20935} & {\it 29982} & 40240 & 50005 & 59468 & {\it 69112} & 80042 & {\it 89808} & 96467 & 105860 & 117317 \\
{\it 5881} & {\it 12062} & {\it 21123} & {\it 30277} & {\it 40326} & 50801 & 59551 &{\it  69879} & 80166 & {\it 89937} & 97150 & {\it 105969} & 117607 \\
{\it 6867} & 12472 & 21433 & 30338 & 41784 & 50966 & 59609 & 69929 & {\it 80346} & {\it 90098} & 97446 & 106267 & \\
{\it 7078} & {\it 12709} & 21673 & 30501 & 42327 & {\it 51157} & 59856 & {\it 72848} & {\it 80686} & {\it 90135} & 97456 & 107136 & \\
7145 & 12716 & 21727 & {\it 31205} & 42542 & {\it 52085} & {\it 60061} & 72939 & {\it 80816} & {\it 90659} & 97594 & 108473 & \\
7487 & {\it 12719} & 22407 & 31646 & 42673 & 52271 & 60129 & {\it 73199} & {\it 81023} & {\it 91751} & 98039 & 109067 & \\
\noalign{\smallskip}
\hline
\end{tabular}
\end{table*}

\begin{figure}[htb]
\resizebox{\hsize}{!}{\includegraphics{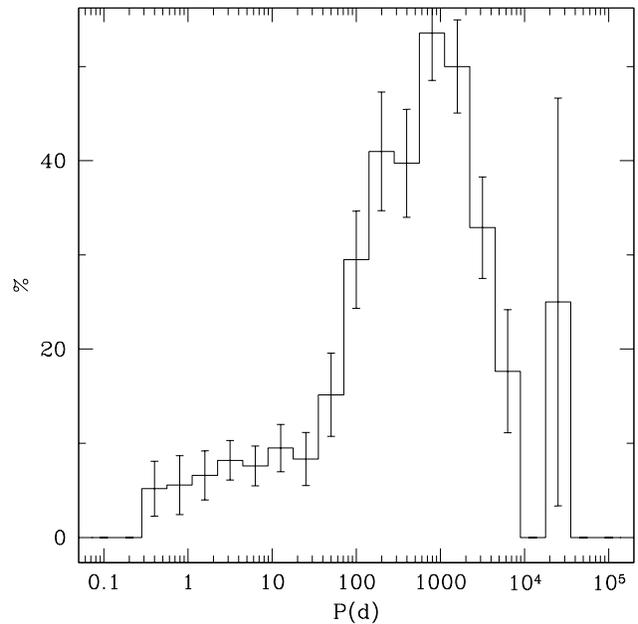}}
\caption[]{\label{Fig:Per_histo1} Percentage of astrometric binaries detected among \SB9\ stars as
a function of orbital period. The error bars give the binomial error on each bin.}
\end{figure}

\begin{figure*}[htb]
\resizebox{0.33\hsize}{!}{\includegraphics{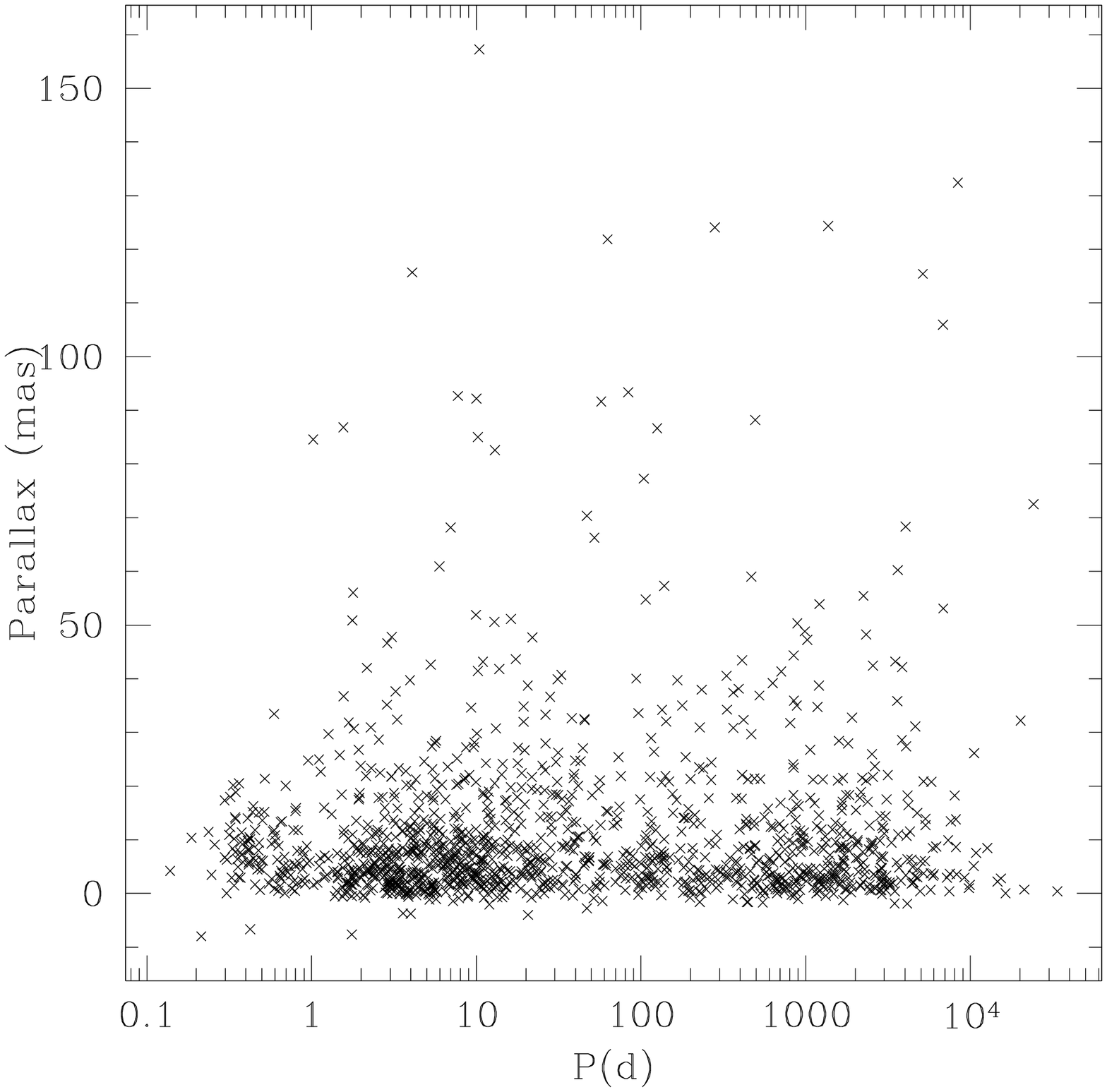}}
\hfill
\resizebox{0.33\hsize}{!}{\includegraphics{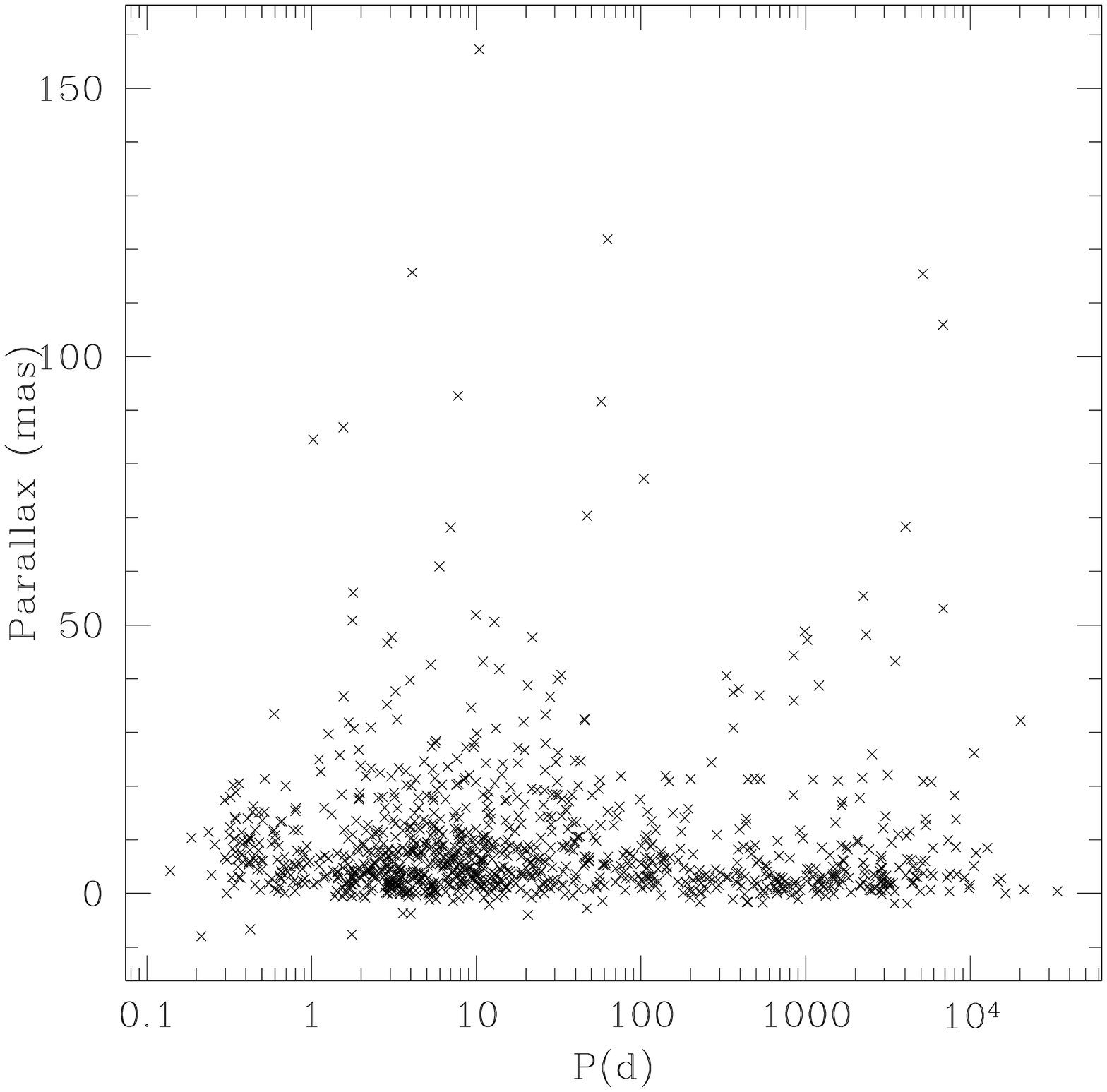}}
\hfill
\resizebox{0.33\hsize}{!}{\includegraphics{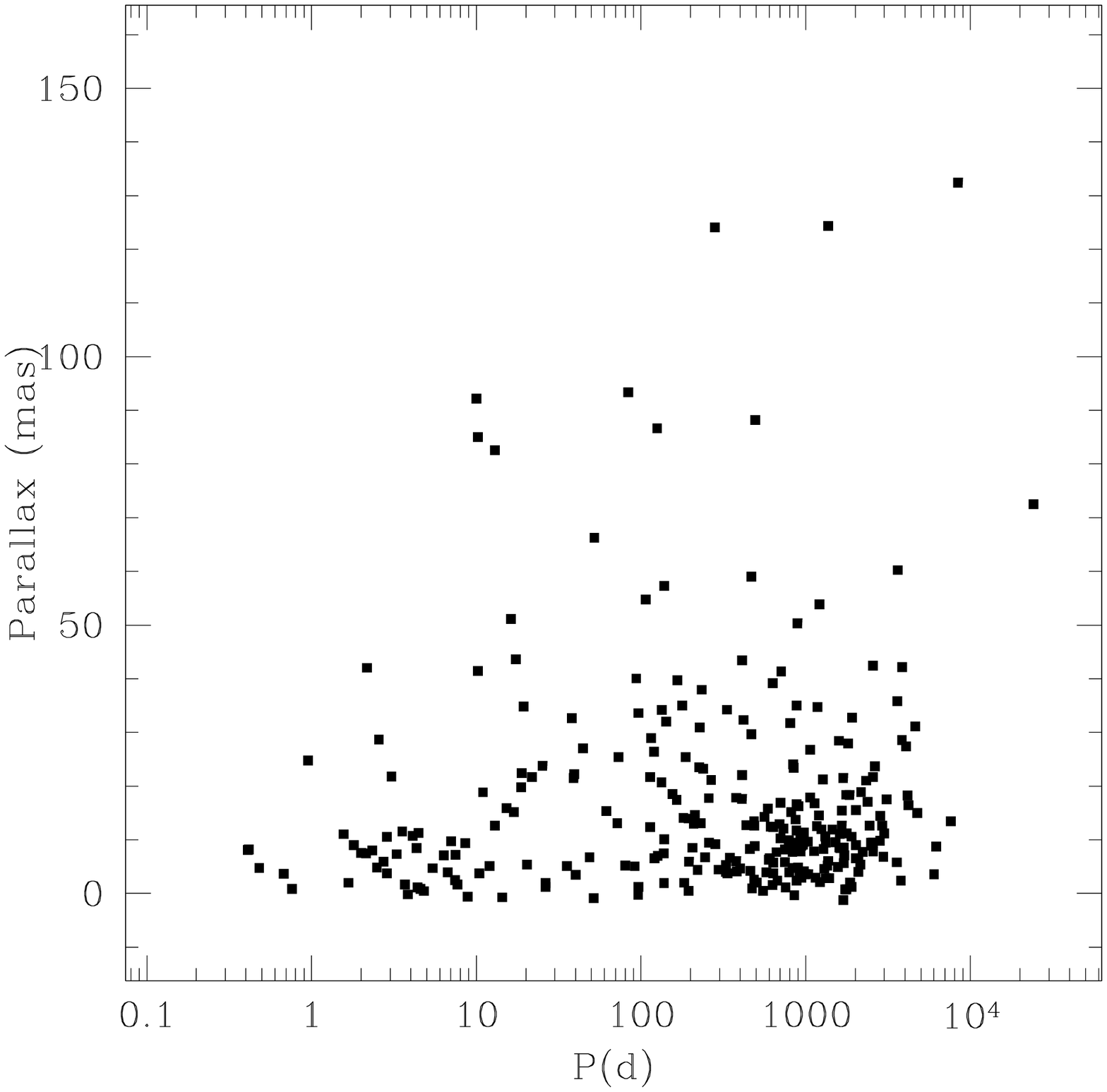}}
\caption[]{\label{Fig:Per_histo2}
{\bf Left panel:} Period-parallax diagram for the selected \SB9 objects with an HIP entry.
{\bf Middle panel:} Period-parallax diagram for non-detected objects.
{\bf Right panel: } Stars flagged as astrometric binaries by the $Pr_1,
Pr_2$ and $Pr_3$ tests at the 5\% level and with $F2_{\mbox{TI}}< 2.37$.}
\end{figure*}

\begin{figure}[htb]
\resizebox{\hsize}{!}{\includegraphics{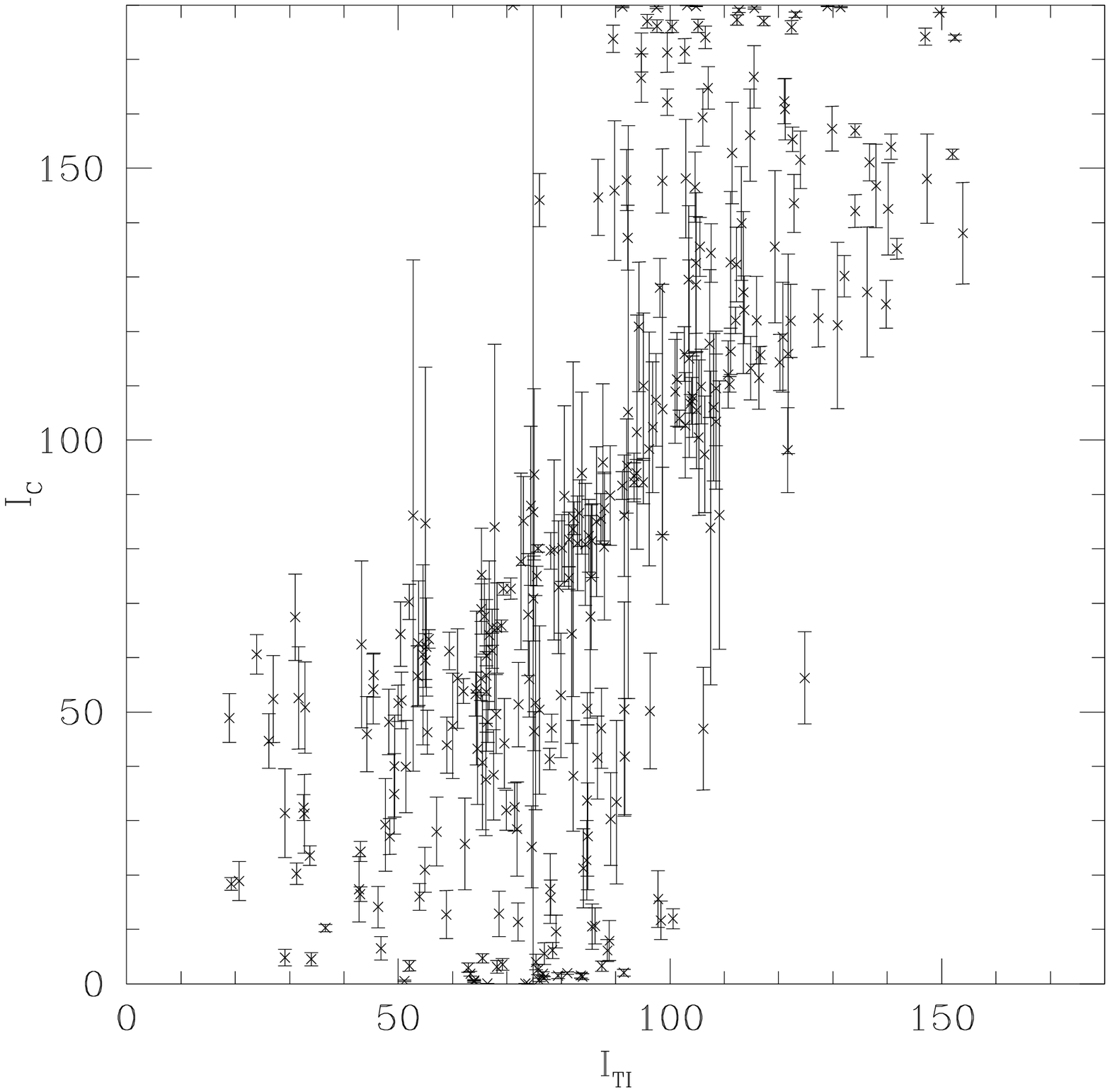}}
\resizebox{\hsize}{!}{\includegraphics{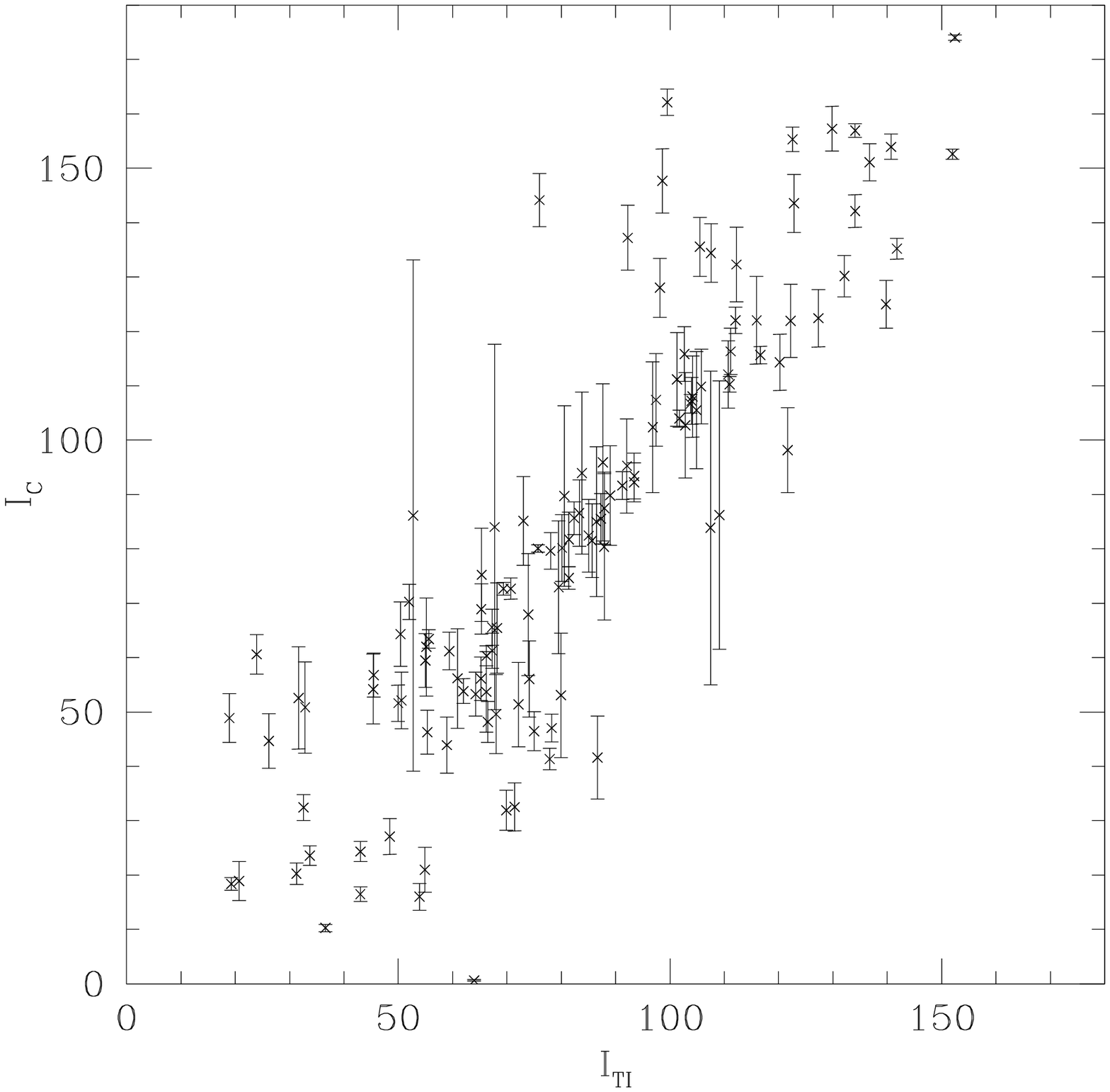}}
\caption[]{\label{Fig:incl_prodex} Comparison of the orbital inclinations derived by the
Thiele-Innes and Campbell approaches. The 282 stars displayed in the upper panel all comply with the 4
criteria for astrometric wobble detection (namely $Pr_1, Pr_2$ and $Pr_3 < 0.05$ and $F2_{\mbox{TI}}< 2.37$ ; see text), but
their astrometric orbital elements are not always reliably determined as not all points fall along
the diagonal. The lower panel displays the 122 stars complying with the $Pr_1, Pr_2$ and $Pr_3$ tests 
at the 0.006\% level.}
\end{figure}

We present the detection rate as a function of the parallax
$\varpi$ and the orbital period $P$ in Table~\ref{Tab:detection} and Figs.~\ref{Fig:Per_histo1} 
and \ref{Fig:Per_histo2}.
A striking property of the astrometric-binary detection rate
displayed in Fig.~\ref{Fig:Per_histo1} is its increase around $P = 50$~d, due to the
Hipparcos scanning law which does not favour the detection of shorter-period binaries.
Similarly, the detection rate drops markedly for periods larger than 2000~d, corresponding to twice
the
duration of the Hipparcos mission. Worth noting are therefore the 5 astrometric orbits detected with
periods larger than 5\ts000~d: HIP~116727 ($P = 24\ts135$~d), HIP~5336 ($P = 8\ts393$~d), HIP~7719 ($P
= 7\,581$~d), HIP~11380 ($P = 6\ts194$~d) and HIP~33420 ($P = 6\ts007$~d). The reason why the
astrometric motion of HIP~116727 could be detected despite such long an orbital period, is that
Hipparcos caught it close to periastron ($e = 0.39$), when the orbital motion is the fastest.
Table~\ref{Tab:detection} further reveals that the detection
rate exhibits little sensitivity to the parallax (provided it is larger than 5~mas; otherwise, the IAD are not precise enough
to extract the
orbital motion), but rather that it is the orbital period which plays the most
significant role. The detection rate in the most favourable cases lies in the range 50 to 80\%. It must be stressed, however, that
{\it all the undetected astrometric binaries in those bins are either SB2
systems, systems with a composite spectrum or with a spectroscopic orbit of
poor quality} (the SB2 and composite-spectrum systems have components of similar
brightness, so that in most cases, the photocenter of the system does not differ much from its
barycenter, making the orbital motion difficult to detect; see Eq.~\ref{Eq:kappabeta} below). If
we remove these entries from the sample, the detection rate is close to 100\%. The orbital
parameters of the detected binaries are further analyzed in Sect.~\ref{Sect:orbit}. Such an analysis
is made necessary when one realizes that the orbital inclinations derived by the Thiele-Innes and
Campbell approaches do not always yield consistent values (Fig.~\ref{Fig:incl_prodex}), contrary to
expectations. Sect.~\ref{Sect:i} therefore presents further criteria used to evaluate the
reliability of the derived astrometric orbital elements (and, in particular, the consistency
between the two sets of orbital parameters, Thiele-Innes versus Campbell).

\begin{table*}[hbt]
\begin{center}
\caption[]{\label{Tab:detection} Detection rate (expressed in \%) as a function of orbital period
and parallax. The percentage is given along with its binomial error; the total number of stars
in the bin is listed between parentheses. For $\varpi > 5$~mas and $100 \le P(\rm d) \le 3000$, the detection rate comes
close to 100\% when removing
SB2 systems, systems with composite spectra or with a poor-quality spectroscopic orbit.}
\setlength{\tabcolsep}{1.9mm}
Period range (d)\\
\vspace{0.3cm}
Parallax (mas)
\hspace{0.2cm}
\begin{tabular}{|l|c|c|c|c|c|}
%\multicolumn{6}{}{Period range}\\
%\hline
%\hline
%&&Period range (d) &&&\\
\hline
& 0--100 & 100--2\ts000 & 2\ts000--3\ts000 & 3\ts000--5\ts000 &   any $P$\\ \hline
 over 15& 16$\pm$3 (177) & 69$\pm$ 5 (81) & 58$\pm$ 14 (12) & 9$\pm$ 12 (12)& 35$\pm$ 3 (292)\\
 10--15 & 7$\pm$ 2 (106) & 69$\pm$ 7 (42)& 80$\pm$ 18 (5) & 18 $\pm$ 19 (5)& 26$\pm$ 3 (162)\\
5-- 10& 8$\pm$ 2 (232) & 51$\pm$ 6 (74)& 65$\pm$ 12 (17) & 18 $\pm$ 14 (5)& 21$\pm$ 2 (336)\\
 0--5 & 7$\pm$ 1 (351) & 24$\pm$ 3 (170) & 4$\pm$ 4 (25)& 5$\pm$ 5 (21)& 12$\pm$ 1 (584)\\
any $\varpi$& 9$\pm$ 1 (866) & 45$\pm$ 3 (367) & 39$\pm$ 6 (59) & 28$\pm$ 7 (43)& 21$\pm$ 1 (1\ts374)\\
\hline
\end{tabular}
\end{center}
\end{table*}

\subsection{The DMSA/O entries}
\label{Sect:DMSAO}
  
Among the 1\ts374 binaries from \SB9, 122 are flagged as DMSA/O in the 
Hipparcos catalogue. We detect 89 of these (or 75\%) (irrespective of $\varpi$).
 The detection rate climbs to  81.7\% (85/104) for orbital periods longer than   100~d.
It is worth examining why not all DMSA/O solutions were retrieved by our processing. 
A close look at the rejected systems reveals that there is nothing wrong with our analysis, since
all but one among the 33 DMSA/O systems not recovered by our reprocessing belong to one of the
following categories:
\begin{itemize}
\item the star is in fact SB2 and possesses an astrometric orbit obtained from
ground-based interferometric or speckle observations; for those cases, 
the DMSA/O solution only provides
$a_0$, with all other parameters taken from the ground-based astrometric solution
(HIP~2912, HIP~10064, HIP~13531, HIP~14328, HIP~14576, HIP~24608, HIP~28360,  HIP~55266, HIP~57565,
HIP~96683, HIP~105431);
\item as indicated in a DMSA/O note, the orbital solution is in fact of poor quality
(`Spectroscopic orbit unreliable. Probably single' {\it sic}), and
does not comply with our more stringent  tests (HIP~10366, HIP~24727, HIP~26563, HIP~35550, HIP~45527);  
\item more orbital elements have been imposed than done here ({\it e.g.,} the inclination, from the
eclipsing nature of the star: HIP~100345; also HIP~23416 = $\epsilon$~Aur);
\item the period provided by the DMSA/O solution is totally different from the one listed in \SB9\
(HIP~8882, HIP~17847, HIP~63613, HIP~82020, HIP~91009 = BY~Dra), sometimes because the system is a
triple one (HIP~85333, HIP~100345);
\item the solution has been marginally rejected by our tests, {\it i.e.,} $Pr_1, Pr_2$ or $Pr_3$ are
only slightly larger than 5\%  (HIP~5778, HIP~32578,  HIP~68756) or $F2$ is slightly larger than 2.37 
(HIP~8833, HIP~59750), or similarly, the DMSA/O solution is
of too low a quality to comply with the tests devised in the present paper (HIP~10324, HIP~12623,
HIP~21273).
\end{itemize}

HIP~85749 is the only DMSA/O solution not belonging to any of the above categories. 
HIP~85749 has not been flagged as an
astrometric binary by our reprocessing, because $Pr_1 = 0.26$, although $Pr_2$, $Pr_3$ and $F2$ 
do qualify
the star as an astrometric binary.
%The astrometric orbit of HIP~45527 has not been retrieved by our processing, probably because the 
%adopted time of periastron passage, extrapolated from the 1928 orbit, has a large cumulative error.
%The DMSA/O solution has left this parameter free instead. 

%\begin{figure}[htb]
%\resizebox{\hsize}{!}{\includegraphics{incl_DMSAO.eps}}
%\caption[]{\label{}Same as Fig.~\ref{Fig:incl_prodex} for the DMSA/O
%objects passing the $Pr_1, Pr_2$ and $Pr_3$ tests.}
%\end{figure}

%
%------------------------------------------------------------------------------
\section{Orbit assessment} \label{Sect:orbit}
%------------------------------------------------------------------------------
%

% \subsection{Consistency between the Thiele-Innes and Campbell solutions}
\label{Sect:i}

The upper panel of Fig.~\ref{Fig:incl_prodex} reveals that, even though orbital solutions pass the $Pr_1, Pr_2$
and $Pr_3$ tests, meaning that an astrometric orbital motion has been detected, these solutions do
not necessarily yield Thiele-Innes and Campbell orbital elements that are consistent with each
other. The inverse-S shape observed  in Fig.~\ref{Fig:incl_prodex} results from the following
properties: (i) in the absence of an orbital signal in the IAD and when the spectroscopic
radial-velocity semi-amplitude
$K_1$ is small, the Campbell solution tends to have
$i_C \sim 0^\circ$ or 180$^\circ$, while the Thiele-Innes solution tends to $i_{TI} \sim 90^\circ$
\citep{Pourbaix-2004:c}; (ii) the physical solutions fall on the diagonal, although this diagonal is
polluted with unphysical solutions having $i_{TI} \sim 90^\circ$.  
The lower panel of Fig.~\ref{Fig:incl_prodex} displays the 122 stars complying with the $Pr_1, Pr_2$ and $Pr_3$ tests 
at the 0.006\% level. It clearly shows that the consistency between the Thiele-Innes and Campbell solutions may be improved considerably by decreasing the probability threshold to 0.006\%.

To remove the remaining inconsistent
solutions, it is necessary to assess the reliability of the derived orbital elements.
This may be done in at least two ways:
\begin{itemize}
\item The consistency between the Thiele-Innes and Campbell elements could be checked 
directly by computing error ellipsoids around the two sets of orbital elements
and estimating whether or not they intersect. This method has not been applied here, because 
it is very time-consuming.

\item Empirical tests have been devised which
check that (i) the astrometric orbital signal has the same period as the adopted spectroscopic
period; (ii) the astrometric orbital
elements should not be too much correlated with each other. This test was already used by \citet
{Pourbaix-2003:a} in a similar context. 
\end{itemize}

The empirical approach has been preferred here, with the two tests 
involved now described in turn.

First, the consistency between the astrometric period and the adopted spectroscopic period is
checked through a periodogram-like test. 
For 600 periods uniformly distributed in $\log P$ between 0.1 and 1200~d, the best
9-parameter (Thiele-Innes) fit is computed (the eccentricity and periastron time are kept
unchanged).  The resulting $\chi^2$ is plotted against the period, thus generating a Scargle-like
periodogram \citep{Scargle-1982:a}. Its
standard deviation $\sigma$ is computed. An orbital motion with the
expected (spectroscopic) period is then supposed to be present in the IAD if the $\chi^2$ at that
period is smaller than the periodogram mean value by more than $\xi \sigma$, with $\xi$ chosen of the
order of 3.

%Second, the expected angular size of the semi-major axis projected on the sky is estimated from the
%spectroscopic orbital elements, namely
%\begin{equation}
%\label{Eq:a1}
%a_1 \sin i =  \varpi  \; \frac{K_1\; P\; \sqrt{1-e^2}} {2 \pi},
%\end{equation}
%and $a_1 = a_0$ if there is no light coming from the companion. This quantity should be compared to
%the standard deviation of the abscissa residuals.

Second, the correlation existing between the Thiele-Innes orbital elements may be estimated through the {\it
efficiency} parameter $\epsilon$ \citep{Eichhorn-1989}, expressed by 
\begin{equation}
\epsilon = \sqrt[p]{\frac{\Pi_{k=1}^p \lambda_k}{\Pi_{k=1}^p \vec{V}_{kk}}},
\end{equation}
where $\lambda_k$ and $\vec{V}_{kk}$ are respectively the eigenvalues and the 
diagonal terms of the covariance matrix $\vec{V}$ and $p$ denotes the number of parameters
in the model.
For an orbital solution to be reliable, its covariance matrix 
should be dominated by the diagonal terms, and the {\it efficiency}
$\epsilon$ should then be close to 1 \citep{Eichhorn-1989}.

\begin{table*}
\caption[]{\label{Tab:NewOrb} The 70 orbital solutions (Campbell solutions)
passing all consistency 
tests. The column
labelled `Ref.' provides the reference for  the spectroscopic orbit used. In the case
where a system is listed in the DMSA/O annex, the column labelled `DMSA' compares the orbital semi-major axes and the inclinations
from the DMSA/O annex and from this work. 
}
\begin{tabular}{rclcccccll}
\hline
\noalign{\smallskip}
HIP    &   $a_0$   & $e$ &  $i$  &   $\omega_1$ & $\Omega$   &  $T_0$   & $P$ & DMSA & Ref. \\
 &  (mas)    &     & (\degr) & (\degr) & (\degr)   & (JD - & (d) & ($a/a_{hip};$ \\
 & & & & & &2\ts400\ts000)& & $i/i_{hip}$) \\
\hline
\noalign{\smallskip}
677 & 7.26$\pm$0.38 & 0.53 & 102.7$\pm$9.7 & 77.5 & 103.4$\pm$5.8 & 47374.6 & 96.7 & O(1.12;0.97) &\citet{Pourbaix-2000}\\
1349 & 20.98$\pm$0.56 & 0.57 & 74.7$\pm$2.1 & 4.7 & 352.3$\pm$3.3 & 34233.3 & 411.4 & O(1.05;0.93) & \citet{Bopp-1970}\\
1955 & 4.68$\pm$0.25 & 0.33 & 108.0$\pm$7.5 & 18.7 & 299.0$\pm$8.3 & 35627.6 & 115.3 & 5  & \citet{Barker-1967}\\
3504 & 7.4$\pm$1.3 & 0.11 & 107.2$\pm$4.3 & 79.0 & 274.5$\pm$4.7 & 41665.0 & 1033.0 & O(1.04;1.06)  &\citet{Abt-1978}\\
6867 & 5.57$\pm$0.46 & 0.00 & 46.3$\pm$4.0 & 0.0 & 340.2$\pm$5.2 & 19544.9 & 193.8 & O(1.13;0.92) &\citet{Luyten-1936}\\
\\
7078 & 7.95$\pm$0.15 & 0.31 & 85.6$\pm$4.7 & 188.2 & 160.5$\pm$3.7 & 29000.4 & 134.1 & O(1.28;0.97) &\citet{Wright-1954}\\
8903 & 12.5$\pm$1.2 & 0.88 & 44.7$\pm$5.0 & 24.9 & 77.8$\pm$5.6 & 44809.1 & 107.0 & O(1.10;1.00) &\citet{Pourbaix-2000}\\
8922 & 8.14$\pm$0.93 & 0.00 & 23.6$\pm$1.8 & 0.0 & 155.7$\pm$4.1 & 43521.5 & 838.0 & X  &\citet{Griffin-1981}\\
10514 & 5.1$\pm$1.4 & 0.06 & 68$\pm$11 & 63.0 & 318$\pm$14 & 41981.5 & 1385.0 & O(0.95;0.92) &\citet{Griffin-1977}\\
11231 & 8.09$\pm$0.34 & 0.29 & 60.6$\pm$3.7 & 188.2 & 200.2$\pm$3.8 & 37159.1 & 142.3 & O(1.43;2.43) &\citet{Barker-1967} \\
\\
12062 & 10.99$\pm$0.87 & 0.26 & 56.8$\pm$4.0 & 254.6 & 64.9$\pm$5.8 & 46440.0 & 905.0 & X  &\citet{Latham-2002} \\
%17932 & 10.5$\pm$1.0 & 0.72 & 92.3$\pm$4.0 & 108.8 & 237.7$\pm$5.8 & 42288.1 & 962.8 & O(1.41;0.99) &\citet{Pedoussaut-1987} \\
20935 & 11.5$\pm$1.1 & 0.24 & 16.5$\pm$1.3 & 127.0 & 308.3$\pm$3.5 & 43298.5 & 238.9 & O(1.06;0.83) & \citet{Griffin-1985:b}\\
24419 & 10.77$\pm$0.59 & 0.08 & 51.6$\pm$3.3 & 275.0 & 230.7$\pm$3.9 & 50690.0 & 803.5 & 9  & \citet{Nidever-2002} \\
26001 & 5.55$\pm$0.46 & 0.51 & 52.2$\pm$5.2 & 330.0 & 45.4$\pm$6.6 & 23108.4 & 180.9 & O(1.21;1.17) &\citet{Lunt-1924}\\

30277 & 9.02$\pm$0.52 & 0.70 & 116.3$\pm$4.2 & 117.1 & 294.6$\pm$5.0 & 19915.0 & 868.8 & O(0.94;1.01) &\citet{Jones-1928}\\
\\
32768 & 7.15$\pm$0.25 & 0.09 & 80.2$\pm$6.1 & 64.0 & 2.9$\pm$6.2 & 20992.8 & 1066.0 & O(0.90;1.00) &\citet{Jones-1928:b}\\
34164 & 8.77$\pm$0.96 & 0.27 & 107.4$\pm$8.5 & 248.9 & 224.3$\pm$8.0 & 47859.9 & 612.3 & X &\citet{Latham-2002}\\
34608 & 4.92$\pm$0.31 & 0.40 & 64.3$\pm$5.9 & 103.4 & 85.6$\pm$6.9 & 44525.0 & 113.3 & O(1.16;0.72) &\citet{Beavers-1985} \\
36377 & 8.32$\pm$0.32 & 0.17 & 65.6$\pm$3.3 & 349.3 & 0.0$\pm$5.2 & 20418.6 & 257.8 & O(1.02;0.96) &\citet{Wilson-1918}\\

39893 & 9.9$\pm$1.3 & 0.21 & 155.3$\pm$2.2 & 210.0 & 193.1$\pm$6.6 & 48342.0 & 733.5 & X  &\citet{Latham-2002}\\
\\
40326 & 10.66$\pm$0.73 & 0.40 & 135.2$\pm$1.9 & 140.0 & 181.7$\pm$2.7 & 18060.0 & 930.0 & O(0.98;0.99) &\citet{Christie-1936}\\
45075 & 10.35$\pm$0.42 & 0.48 & 61.3$\pm$3.2 & 349.4 & 119.2$\pm$3.8 & 25721.6 & 1062.4 & O(0.92;0.91) &\citet{Bretz-1961} \\
47461 & 4.19$\pm$0.67 & 0.15 & 122.0$\pm$8.0 & 135.4 & 282.7$\pm$9.7 & 45464.5 & 635.4 & 5  &\citet{Ginestet-1991}\\
52085 & 8.09$\pm$0.62 & 0.10 & 125.0$\pm$4.4 & 270.0 & 327.5$\pm$6.9 & 20760.0 & 1200.0 & O(1.36;0.85) &\citet{Christie-1936}\\
53240 & 7.83$\pm$0.93 & 0.38 & 134.4$\pm$5.4 & 301.0 & 292.4$\pm$6.2 & 42901.5 & 1166.0 & O(0.90;1.09) &\citet{Griffin-1980}\\
\\
57791 & 7.47$\pm$ 0.60 & 0.31 & 86.5 $\pm$ 6.1 & 125.1 & 108.5$\pm$ 4.8& 42352.7& 486.7& O(0.97;0.99) & \citet{Ginestet-1985}\\
%59468 & 3.98$\pm$ 0.88&  0.17& 47.0$\pm$ 7.3& 235.3&  270.8$\pm$8.8 &22360.8 &461. & O(0.82;0.68) &\citet{Harper-1930}\\
%59750 & 28.24$\pm$0.56 & 0.05 & 75.0$\pm$1.8 & 199.0 & 193.0$\pm$2.1 & 47589.0 & 843.9 & O(0.97;1.01) & n &\citet{Carney-2001}\\
60998 & 5.99$\pm$0.93 & 0.30 & 32.5$\pm$4.4 & 244.0 & 205.7$\pm$8.6 & 42868.0 & 1703.0 & 7  &\citet{Reimers-1988}\\
61724 & 8.40$\pm$0.82 & 0.59 & 81.5$\pm$6.8 & 102.5 & 139.0$\pm$7.0 & 43304.0 & 972.4 & O(0.84;0.96) &\citet{Griffin-1981:b}\\
62915 & 5.35$\pm$0.76 & 0.32 & 27.1$\pm$3.3 & 194.0 & 40.7$\pm$ 8.4 & 43424.5 & 1027. & 9  &\citet{Griffin-1983} \\
63406 & 14.12$\pm$0.47 & 0.33 & 81.8$\pm$5.0 & 65.0 & 101.3$\pm$4.9 & 49220.0 & 710.6 & O(0.86;1.01) & \citet{Griffin-2002:a}\\
\\
65417 & 12.0$\pm$1.7 & 0.19 & 68.9$\pm$4.6 & 166.0 & 121.1$\pm$3.7 & 45497.5 & 1366.8 & O(1.17;1.13) &\citet{Griffin-1986}\\
67234 & 6.45$\pm$0.61 & 0.13 & 48.2$\pm$3.8 & 58.6 & 280.3$\pm$4.9 & 24163.0 & 437.0 & O(1.01;0.77) &\citet{Jones-1928}\\
67927 & 36.02$\pm$0.56 & 0.26 & 115.7$\pm$1.6 & 326.3 & 75.2$\pm$1.4 & 28136.2 & 494.2 & O(1.02;0.99) &\citet{Bertiau-1957}\\
67480 & 7.3 $\pm$ 0.9 &  0.41 & 174.0 $\pm$ 0.5 &  359   &278.8 $\pm$ 5.8 &   44739.5  &  944&X &\citet{Griffin-1985} \\
69112 & 5.74$\pm$0.55 & 0.14 & 130.2$\pm$3.8 & 311.8 & 158.5$\pm$5.3 & 38901.7 & 605.8 & O(1.00;0.94) &\citet{Scarfe-1971}\\
\\
69879 & 4.65$\pm$0.24 & 0.57 & 89.8$\pm$9.1 & 224.9 & 347.0$\pm$8.6 & 40286.0 & 212.1 & O(1.26;1.02) &\citet{Scarfe-1975}\\
72848 & 16.54$\pm$0.18 & 0.51 & 93.4$\pm$4.2 & 219.0 & 248.3$\pm$3.6 & 50203.4 & 125.4 & O(1.16;0.94) &\citet{Nidever-2002}\\
73199 & 6.10$\pm$0.48 & 0.13 & 53.3$\pm$4.0 & 212.0 & 239.3$\pm$5.5 & 44419.0 & 748.9 & O(0.85;0.84) &\citet{Batten-1986}\\
73440 & 4.69$\pm$0.47 & 0.22 & 43.9$\pm$5.2 & 10.0 & 288.5$\pm$6.5 & 47349.0 & 467.2 & X  &\citet{Latham-2002}\\
74087 & 11.2$\pm$1.6 & 0.83 & 62.0$\pm$9.0 & 175.3 & 82.6$\pm$6.6 & 48356.6 & 2567.1 & 7  &\citet{Griffin-1999}\\
\\
75379 & 8.4$\pm$1.3 & 0.68 & 52.6$\pm$9.4 & 339.5 & 215.5$\pm$7.5 & 14785.1 & 226.9 & O(1.00;1.07) &\citet{Jones-1931}\\
%77409 & 6.1$\pm$1.1 & 0.83 & 118$\pm$14 & 339.7 & 299$\pm$15 & 50926.3 & 233.1 & 5 &\citet{Griffin-2003}\\
79101 & 8.65$\pm$0.64 & 0.47 & 10.3$\pm$0.7 & 357.0  &  148.3$\pm$3.0 & 40525.2 & 560.5 & O &\citet{Aikman-1976}\\
80346 & 53.0$\pm$1.8 & 0.67 & 152.60$\pm$0.95 & 251.0 & 286.9$\pm$1.5 & 51298.0 & 1366.1 & O(1.02;1.03)  &\citet{Nidever-2002} \\
80686 & 2.71$\pm$ 0.44 & 0.06 & 16.01 $\pm$ 2.47 & 274.5  & 2.1$\pm$6.8 & 18103.6 & 12.9 & 5 &\citet{Jones-1928}\\
80816 & 11.37$\pm$0.51 & 0.55 & 53.8$\pm$2.3 & 24.6 & 341.9$\pm$3.8 & 15500.4 & 410.6 & O(1.03;1.16)& \citet{Plummer-1908}\\
\noalign{\smallskip}
\hline
\end{tabular}
\end{table*}

\setlength{\tabcolsep}{5pt}
\addtocounter{table}{-1}
\begin{table*}
\caption[]{Continued.}
\begin{tabular}{rclcccccll}
\hline
\noalign{\smallskip}
HIP    &   $a_0$   & $e$ &  $i$  &   $\omega_1$ & $\Omega$   &  $T_0$   & $P$ & DMSA & Ref. \\
 &  (mas)    &     & (\degr) & (\degr) & (\degr)   & (JD - & (d) & ($a/a_{hip};$ \\
 & & & & & &2\ts400\ts000)& & $i/i_{hip}$) \\
\hline
\noalign{\smallskip}
82860 & 6.58$\pm$0.32 & 0.21 & 56.1$\pm$4.0 & 339.0 & 228.7$\pm$4.5 & 39983.6 & 52.1 & O(0.98;0.90) &\citet{Abt-1976}\\
83575 & 9.11$\pm$0.58 & 0.22 & 61.2$\pm$3.5 & 348.0 & 19.5$\pm$4.3 & 46806.0 & 790.6 & O(1.04;1.03) &\citet{Griffin-1991}\\
86400 & 12.71$\pm$0.78 & 0.23 & 18.4$\pm$1.1 & 140.5 & 274.2$\pm$2.7 & 47724.9 & 83.7 & O(0.92;0.41) & \citet{Tokovinin-1991} \\
87895 & 29.81$\pm$0.62 & 0.41 & 72.7$\pm$1.2 & 134.8 & 177.4$\pm$1.0 & 47714.6 & 881.8 & O(1.09;1.07) & \citet{Pourbaix-2000} \\
88788 & 10.09$\pm$ 1.04 & 0.378 & 153.9$\pm$ 2.3 & 137.0 & 341.4$\pm$5.8 & 46139.0 & 2017.0 & 9 &\citet{Griffin-1992}\\
\\
89937 & 50.30$\pm$0.23 & 0.41 & 80.08$\pm$0.70 & 119.9 & 232.42$\pm$0.83 & 46005.6 & 280.5 & O(1.25;1.07) &\citet{Pourbaix-2000} \\
90659 & 9.1$\pm$1.1 & 0.50 & 142.1$\pm$3.0 & 56.0 & 353.1$\pm$6.0 & 42925.5 & 1284.0 & O(0.97;1.01) &\citet{Griffin-1980}\\
91751 & 6.41$\pm$0.53 & 0.21 & 59.5$\pm$4.9 & 78.0 & 295.8$\pm$7.0 & 42928.5 & 485.3 & O(1.05;1.05) &\citet{Griffin-1982}\\
92512 & 3.16$\pm$0.25 & 0.11 & 106$\pm$11 & 274.3 & 0$\pm$37 & 19258.2 & 138.4 & O(1.02;1.09) &\citet{Young-1920}\\
93244 & 12.80$\pm$0.44 & 0.27 & 87.5$\pm$6.6 & 82.0 & 58.7$\pm$3.9 & 41718.5 & 1270.6 & O(0.94;1.00)  &\citet{Griffin-1982}\\
\\
95028 & 3.0 $\pm$ 0.7 &  0.37 & 19 $\pm$ 4 &  161  &242.9 $\pm$ 10.0 & 43811.9  &  208.8&7 &\citet{Griffin-1982}\\
95066 & 9.15$\pm$0.54 & 0.83 & 75.2$\pm$8.6 & 152.7 & 129.7$\pm$6.7 & 33420.2 & 266.5 & O(1.18-1.05) &\citet{Franklin-1952}\\
95575 & 8.14$\pm$0.25 & 0.15 & 98.2$\pm$7.8 & 63.3 & 208.1$\pm$5.7 & 47746.4 & 166.4 & X  &\citet{Tokovinin-1991}\\
99848 & 5.5$\pm$1.2 & 0.30 & 65.5$\pm$8.3 & 218.2 & 0$\pm$227 & 33141.8 & 1147.8 & O(1.04;1.02) &\citet{Wright-1970}\\
99965 & 14.06$\pm$0.39 & 0.08 & 92.2$\pm$3.6 & 243.0 & 303.9$\pm$4.5 & 50218.0 & 418.8 & O(0.87;0.94) &\citet{Griffin-2002:c}\\
\\
100437 & 5.35$\pm$0.57 & 0.76 & 54.2$\pm$6.4 & 108.1 & 249.7$\pm$8.2 & 49281.0 & 1124.1 & 9  & \citet{Griffin-2000:b}\\
101093 & 19.79$\pm$0.55 & 0.03 & 104.0$\pm$1.6 & 83.7 & 95.4$\pm$1.7 & 16214.5 & 840.6 & O(1.39;1.01) &\citet{Abt-1961}\\
101847 & 3.82$\pm$ 0.81 & 0.0 & 147.7$\pm$ 5.9 & 0.0  & 314.0$\pm$1.3 & 23358.0 & 205.2 & 5 &\citet{Lucy-Sweeney-1971}\\
103519 & 7.48$\pm$0.64 & 0.44 & 32.4$\pm$2.4 & 148.1 & 306.2$\pm$4.0 & 39186.1 & 635.1 & O(1.02;0.94) & \citet{Radford-1975}\\
105969 & 10.65$\pm$0.75 & 0.13 & 157.0$\pm$1.3 & 192.0 & 236.6$\pm$3.9 & 47479.0 & 878.0 & X  &\citet{McClure-1997}\\
\\
%107089 & 29.34$\pm$0.49 & 0.40 & 65.9$\pm$1.1 & 80.0 & 116.7$\pm$1.2 & 18525.0 & 1020.0 & X  & n &\citet{Christie-1936}\\
109176 & 4.07$\pm$0.27 & 0.00 & 80$\pm$13 & 0.0 & 188$\pm$11 & 45320.0 & 10.2 & 5 & \citet{Fekel-1983}\\
111170 & 26.67$\pm$0.73 & 0.38 & 63.4$\pm$1.7 & 171.6 & 82.0$\pm$1.9 & 43995.0 & 630.1 & O(1.14;1.07)  & \citet{Pourbaix-2000}\\
112158 & 15.62$\pm$0.85 & 0.15 & 72.7$\pm$1.9 & 5.6 & 208.6$\pm$2.3 & 15288.7 & 818.0 & O(1.15;1.03) &\citet{Crawford-1901}\\
113718 & 12.16$\pm$0.93 & 0.54 & 24.3$\pm$1.8 & 247.7 & 328.6$\pm$5.6 & 48280.0 & 468.1 & O(1.10;0.41) &\citet{Latham-2002}\\
114421 & 7.82$\pm$ 0.47& 0.66& 114.3$\pm$5.2 &240.8& 120.8$\pm$7.2 &16115.6& 409.6& O(1.26;0.93) &\citet{Jones-1928}\\
\noalign{\smallskip}
\hline
\end{tabular}
\end{table*}

\setlength{\tabcolsep}{4.5pt}
\begin{table*}
\caption[]{\label{Tab:orbits_medium} The 31 new orbital solutions (Campbell solutions)
passing the $Pr_1, Pr_2$ and $Pr_3$ tests at the 0.006\% level, but failing at least one of the consistency 
tests. The column
labelled `Ref.' provides the reference for  the spectroscopic orbit used. 
The columns labelled $\xi$ and $\epsilon$  provide the values of the corresponding empirical tests. The column labelled `D' refers to DMSA.}
\begin{tabular}{rclccccccccl}
\hline
\noalign{\smallskip}
HIP & $a_0$   & $e$ &  $i$  &   $\omega_1$ & $\Omega$   &  $T_0$   & \multicolumn{2}{l}{\;\;\;\;$P$  \;\;\;\;\;\;D} & $\xi$ & $\epsilon$ & Ref. \\
 &  (mas)    &     & (\degr) & (\degr) & (\degr)   & (JD - & (d) &  \\
 & & & & & &2\ts400\ts000)&    \\
\hline
\noalign{\smallskip}
2081  & 103.5$\pm$8.2 & 0.34 & 128.0$\pm$5.4& 19.8  & 242.8 $\pm$3.9     &16201.8 & 3848.8   & 7 &4.25&0.01& \citet{Lunt-1924}\\	     
5881 & 7.7$\pm$1.6 & 0.12 & 157.2$\pm$4.1& 313  &    236.8$\pm$ 8.1	&51791.1 & 701.4    & 5 &1.91&0.42& \citet{Nidever-2002}\\	
11349 & 84.4$\pm$4.5 & 0.01 & 115.8$\pm$5.0& 225.  & 44.4$\pm$ 8.2	&45901. & 3600.     & 9 &3.31&0.07& \citet{Latham-2002}\\		
11923 & 33$\pm$32 & 0.54 & 86.2$\pm$24.6& 259.0  &      83$\pm$ 20	&47774.2 & 2332.    & 7 &4.39&0.07& \citet{Latham-2002}\\	
13055 & 8.12$\pm$0.77 & 0.09 & 85.0$\pm$13.7& 120  &   279$\pm$ 16	&46344 & 2018       & 7 &4.41&0.34& \citet{McClure-1990}\\
\\		
15394 & 21.9$\pm$3.0 & 0.86 & 137.2$\pm$5.9& 71.7  &    81$\pm$ 10	&51190.3 & 3089.4   & 7 &2.18&0.11& \citet{Latham-2002}\\	
27246 & 56.3$\pm$6.8 & 0.32 & 49.6$\pm$7.2& 318.5  &344.6$\pm$ 4.4	&49649. & 4072.     & 9 &5.11&0.02& \citet{Latham-2002}\\		
31681 & 78.7$\pm$2.3 & 0.89 & 106.7$\pm$1.7& 312.6 &243.6$\pm$ 2.6	&43999.1 & 4614.5   & X &2.61&0.01& \citet{Lehmann-2002}\\	
38414 & 32.7$\pm$2.2& 0.38 & 41.3$\pm$1.9& 170.  & 148.5$\pm$ 3.6	&17031. & 2554.0    & 9 &5.17&0.10& \citet{Parsons-1983}\\	
39424 & 19.5$\pm$2.9 & 0.06 & 50.8$\pm$8.4& 264.  & 242.6$\pm$ 5.5	&42894.5 & 2437.8   & 7 &2.78&0.11& \citet{Griffin-1982b}\\	
\\
43903 & 37$\pm$3422 & 0.70 & 84$\pm$33& 194.1  &       184$\pm$ 21	&49093.7 & 1898.7   & 7 &5.67&0.05& \citet{Carney-2001}\\	
44946 & 10.5$\pm$1.7 & 0.06 & 144.1$\pm$4.9& 301.1  &282.0$\pm$ 9.7	&28876.8 & 1700.7   & 7 &2.88&0.28& \citet{Jackson-1957}\\	
46893 & 4.69$\pm$0.77 & 0.15 & 132.3$\pm$6.8& 261.  &     2$\pm$ 11	&43119.5 & 830.4    & X &2.84&0.86& \citet{Griffin-1981}\\	
51157 & 17.9$\pm$1.2 & 0.86 & 122.4$\pm$5.2& 296.1  & 255.7$\pm$ 5.5	&44583.0 & 1180.6   & 9 &2.78&0.41& \citet{Griffin-1987}\\	
53238 & 29.6$\pm$5.0 & 0.16 & 143.5$\pm$5.3& 285.  & 221.2$\pm$ 7.0	&45781. & 1841.     & 7 &2.73&0.21& \citet{Latham-2002}\\		
\\
55016 & 19.0$\pm$3.1 & 0.41 & 53.6$\pm$7.3& 336.5  & 287$\pm$ 10	&42054. & 2962.7    & 7 &3.00&0.15& \citet{Wolff-1974}\\	
60061 & 19.6$\pm$2.7 & 0.41 & 51.3$\pm$7.7& 302.6  & 11.8$\pm$ 8.2	&50134. & 2167.     & 7 &3.33&0.28& \citet{Latham-2002}\\		
68072 & 10.8$\pm$2.7 & 0.68 & 20.9$\pm$4.1& 177.6  & 6$\pm$ 10		&47179.1 & 1620.3   & 7 &4.43&0.32& \citet{Latham-2002}\\	
75718 & 38.6$\pm$1.0 & 0.97 & 60.3$\pm$1.8& 253.9  & 95.8$\pm$ 3.5	&47967.5 & 889.6    & 7 &3.20&0.29&\citet{Duquennoy-1992}\\	
79358 & 16.5$\pm$2.3 & 0.6 & 46.4$\pm$3.5& 340.  & 305.4$\pm$ 6.7	&24290. & 2150.     & 7 &4.73&0.18& \citet{Christie-1936}\\		
\\
84949 & 24.0$\pm$1.1 & 0.67 & 70.2$\pm$3.2& 40.0  & 150.4$\pm$ 2.1	&44545.8 & 2018.8   & X &4.31&0.30& \citet{Scarfe-1994}\\	
86722 & 49.4$\pm$8.5 & 0.93 & 41.6$\pm$7.6& 129.6  & 315$\pm$ 11	&49422.5 & 2558.4   & 7 &3.50&0.01& \citet{Duquennoy-1996}\\	
90098 & 30.7$\pm$4.1 & 0.26 & 56.0$\pm$7.0& 187.2  & 54.6$\pm$ 8.4	&18076.2 & 2214.    & 9 &3.23&0.06& \citet{Jones-1928}\\	
90135 & 21.6$\pm$1.9 & 0.10 & 89$\pm$16.5& 242.1  & 226$\pm$ 14		&18278.3 & 2373.7   & 7 &2.85&0.13&\citet{Grobben-1969}\\	
92872 & 26.6$\pm$3.4 & 0.24 & 31.9$\pm$3.6& 35.  & 12.6$\pm$ 7.7	&44276.5 & 2994.    & 7 &3.53&0.04& \citet{Griffin-1981}\\	
\\
94371 & 33.9$\pm$3.8 & 0.19 & 135.5$\pm$5.4& 103.  & 126.9$\pm$ 6.1	&41044.5 & 2561.    & 7 &3.52&0.09& \citet{Griffin-1979}\\	
103987 & 10.7$\pm$2.7 & 0.08 & 162.1$\pm$2.3& 83.  & 15.6$\pm$ 3.9	&46639.0 & 377.8    & 9 &2.34&0.23& \citet{Latham-1992}\\	
114313 & 17.2$\pm$2.2 & 0.22 & 20.2$\pm$1.9& 237.  & 75.1$\pm$ 4.8	&46444. & 1132.     & 9 &3.31&0.38& \citet{Latham-2002}\\		
116478 & 20.9$\pm$1.1 & 0.33 & 109.8$\pm$6.8& 304.3  & 129.1$\pm$ 7.8	&47403. & 1810.     & 9 &4.60&0.16& \citet{Latham-2002}\\		
116727 & 376$\pm$23189 & 0.38 & 86$\pm$47& 166.0  & 16$\pm$ 19		&48625 & 24135      & 7 &5.56&0.01& \citet{Griffin-2002:b}\\	
\\
117229 & 9.38$\pm$0.78 & 0.52 & 102$\pm$12& 192.3  & 251.7$\pm$ 8.4	&48425.8 & 1756.0   & 7 &2.93&0.59& \citet{Latham-2002}\\ 
\noalign{\smallskip}
\hline
\end{tabular}
\end{table*}

\begin{figure}[htb]
\resizebox{\hsize}{!}{\includegraphics{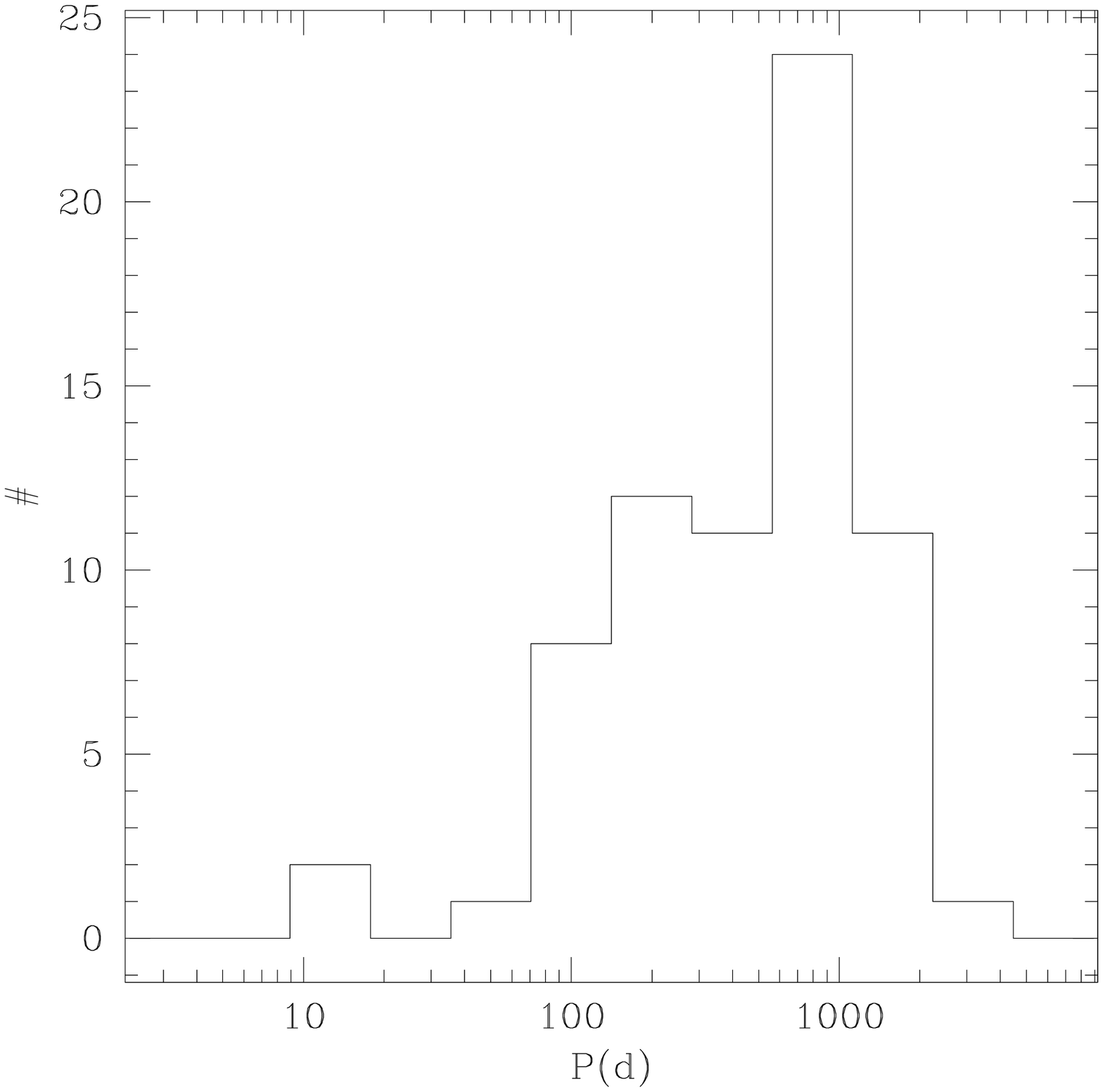}}
\caption{\label{Fig:period} Distribution of the orbital periods for the 70 solutions retained.}
\end{figure}

\begin{figure}[htb]
\resizebox{\hsize}{!}{\includegraphics{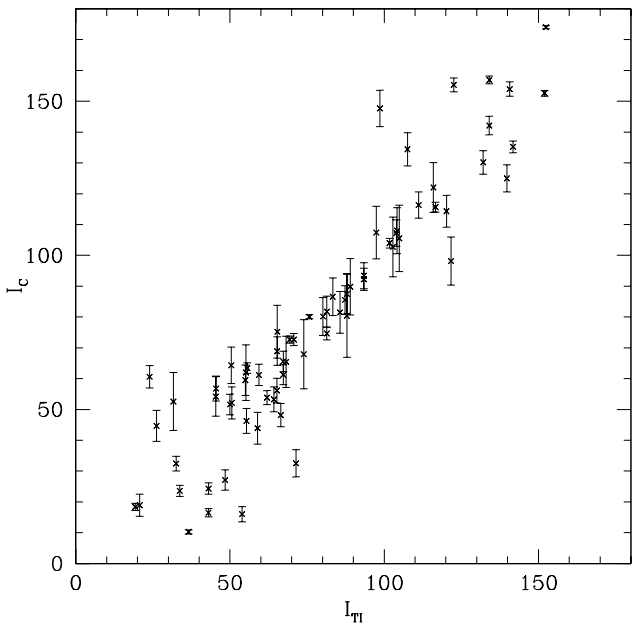}}
\caption[]{\label{Fig:empirical} Comparison of the inclinations derived from the Thiele-Innes
constants and from the Campbell elements for the 70 systems retained. Compare with
Fig.~\protect\ref{Fig:incl_prodex}.}
\end{figure}

The 70 orbital solutions retained when adopting  $\xi = 3$ and
$\epsilon > 0.4$ are  listed in Table~\ref{Tab:NewOrb}, 20 of them being new orbital solutions
not  already listed in the DMSA/O annex. Fig.~\ref{Fig:period} presents the distribution of their orbital
periods. In Fig.~\ref{Fig:empirical} comparing the inclinations derived from the Thiele-Innes and
Campbell solutions, the retained orbits now fall close to the diagonal, as expected.

Neither the parallax nor the proper motions differ significantly  from the Hipparcos value for 
the stars of Table~\ref{Tab:NewOrb}. They have therefore not been listed.

To increase the science content of this paper, Table~\ref{Tab:orbits_medium} lists the astrometric 
orbital elements for a second category of systems: 31 newly derived orbits ({\it i.e.,} 
not already present in the DMSA/O annex), not already listed in Table~\ref{Tab:NewOrb}, 
from the list of 122 stars passing the $Pr_1, Pr_2$ and $Pr_3$ tests at the 0.006\% level (they are among the italicized stars in 
Table~\ref{Tab:Pr123}). These orbits are (possibly) of a slightly lower accuracy as the ones listed in Table~\ref{Tab:NewOrb}
because they do not comply with the two empirical tests described in this section. Nevertheless, 
these newly derived orbits are worth publishing.

\begin{figure}[htb]
\resizebox{\hsize}{!}{\includegraphics{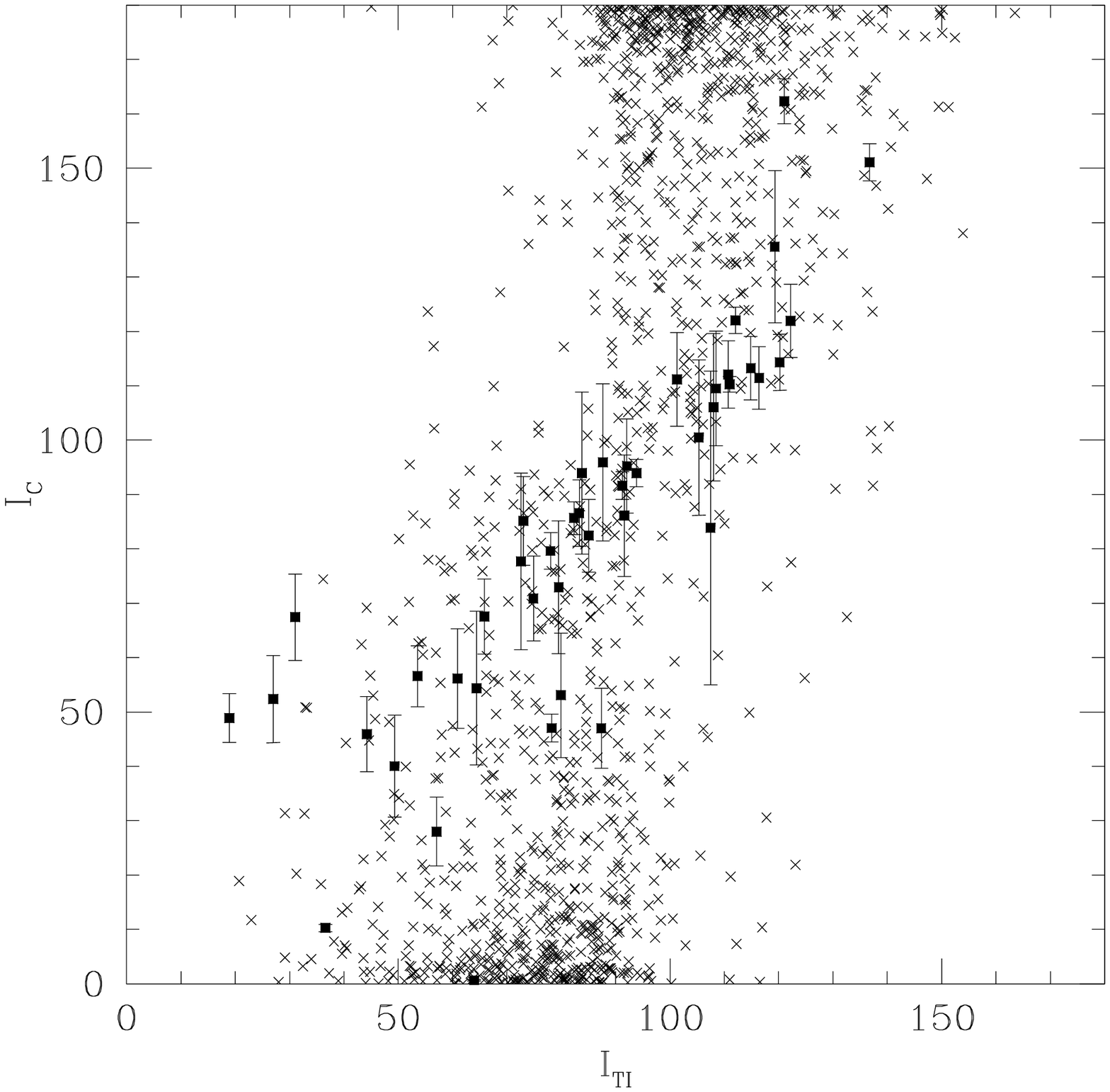}}
\caption[]{\label{Fig:Mag3} Comparison of the inclinations derived from the Thiele-Innes constants
and from the Campbell elements for the 1304 systems not retained. The 39 rejected systems 
with a solution in the
DMSA/O annex (Table~\ref{Tab:DMSAOno}) are represented by a filled square.}
\end{figure}

As already discussed in Sect.~\ref{Sect:DMSAO}, there are 122 systems in our sample of 1\ts 374
which have a DMSA/O entry. Of these 122, 89 pass the $Pr_1, Pr_2$, $Pr_3$ and $F2$ tests at the 5\% level
(Sect.~\ref{Sect:DMSAO}) and 71 pass the $Pr_1, Pr_2$, $Pr_3$ and $F2$ tests at the 0.006\% level  but only 50
have reliable orbital elements according to the 2 empirical
tests described in this section.  The 39 rejected DMSA/O systems are listed in
Table~\ref{Tab:DMSAOno}, along with the failed test(s). 
%In two cases (HIP~56731, HIP~59856), the
%Wilks test is satisfied but not the efficiency test; these systems are nevertheless included in
%Table~\ref{Tab:NewOrb} containing the accepted orbits. 
Fig.~\ref{Fig:Mag3} compares the
Thiele-Innes and Campbell inclinations for those systems with orbital elements not validated by the
consistency tests. 

The orbits derived in the present analysis and the DMSA/O ones generally agree well. For HIP~677
(= $\alpha$~And), a visual and SB2 system, there are  astrometric orbits based on ground-based
interferometric measurements already available \citep{Pan-1992, Pourbaix-2000}. The inclination of
$103^\circ\pm10^\circ$ found here is consistent with the value  $105.7^\circ\pm0.2^\circ$ obtained
by \citet{Pan-1992}. The only new constraint of interest provided by the
IAD-derived photocentric orbit lies in a consistency check between that photocentric semi-major axis
$a_0 = 7.3\pm0.4$~mas (Table~\ref{Tab:NewOrb}) and the relative  semi-major axis $a = 24.1\pm0.1$~mas
\citep{Pan-1992, Pourbaix-2000}, with the following relation to be satisfied \citep{Binnendijk}:
\begin{equation}
\label{Eq:kappabeta}
a_0 = a (\kappa - \beta),
\end{equation}
where $\kappa = M_2 / (M_1 + M_2) = 0.331$
and $\beta = (1 + 10^{0.4 \Delta m})^{-1}$, and 
$\Delta m$
is the magnitude difference between the two components. Eq.~\ref{Eq:kappabeta} then implies
$\beta = 0.027$ or $\Delta m = 3.9$~mag, which is much larger than the value of 2.0~mag  measured 
by \cite{Pan-1992} or 2.19~mag derived by  \citet{Ryabchikova-1999}. With $\Delta m = 2$~mag, $\beta
= 0.137$, so that $a_0/a = 0.19$ or $a_0 = 4.7$~mas, which is inconsistent with the value of
$7.3\pm0.4$~mas listed in Table~\ref{Tab:NewOrb} or $a_0 = 6.47\pm1.16$~mas from the DMSA/O. The origin of this discrepancy is unknown.

\begin{table}
\caption[]{\label{Tab:DMSAOno}
The 39 systems with a DMSA/O entry which do not fulfill the 2 tests assessing the reliability of the
astrometric orbital elements, namely $\xi < 3$ and $\epsilon > 0.4$  and the probability tests at the 0.006\% level (see
text). Columns with `n' correspond to failed tests. }
\begin{tabular}{llllp{4cm}}
\hline
\noalign{\smallskip}
HIP & $\xi$ &  $\epsilon$ & $Pr$ & Rem.\cr
\noalign{\smallskip}
\hline
443 & y & y & n \cr
5336 & y & n & y \cr
%8833 & y & n & y \cr
%10366 & n & y & y \cr
10644 & y & y & n \cr
10723 & n & y & y \cr
12709 & y & n & y \cr
12719 & n & y & y & test failing only marginally \cr
16369 & n & y & n \cr
17296 & n & y & n \cr
17440 & y & n & y \cr
17932 & y & y & n \cr
20070 & n & y & y \cr
20087 & n & n & y & DMSA/O solution provides only $a_0$\cr
20482 & n & y & n \cr
21123 & n & y & y \cr
23453 & n & y & y & test failing only marginally \cr
23922 & n & n & n & tests failing only marginally; DMSA/O solution from scratch providing a period
different from the \SB9\ one\cr 
%24727 & n & y & y \cr
29982 & n & n & y \cr
30501 & n & y & n \cr
31205 & n & y & y &  test failing only marginally \cr
32761 & y & y & n &  test failing only marginally \cr
49841 & y & n & y \cr
52419 & n & y & n & DMSA/O solution from scratch providing a period
different from the \SB9\ one \cr
56731 & y & n & y & test failing only marginally\cr
58590 & n & y & n \cr
59459 &  n & y & n & test failing only marginally \cr
59468 & y & y &n \cr
59856 &  n & n & n & test failing only marginally \cr
68682 & y & n & y \cr
75695 & y & n & y \cr
76267 & n & y & n \cr
%79101 & y & y & n \cr
80166 & n & y & n & tests failing only marginally\cr
81023 & n & y & y \cr
89808 & n & y & y \cr
92175 & n & y & n & test failing only marginally \cr
92818 & y & y & n & test failing only marginally \cr
99675 & y & n & y \cr
109554 & n & y & n & test failing only marginally\cr
110130 & y & n & y \cr
113860 & y & n & y \cr
\noalign{\smallskip}
\hline
\end{tabular}
\end{table}

In Table~\ref{Tab:DMSAOno}, cases where the efficiency test is the only one to fail generally correspond to rather wide orbits
which cannot be accurately determined with Hipparcos data only ({\it e.g.}, HIP~5336, 68682,
75695, 110130). 
%Cases where only the $a_1 \sin i$ test fails correspond instead to orbits
%with semi-major axes too small to be accurately determined. 
When only the periodogram test
fails, it means that either the spectroscopic period does not correspond to the astrometric motion,
or that the IAD  do not constrain its period well enough.

%We reject the null hypothesis ``{\em Thieles-Innes and Campbell's approaches yield consistent
%solutions}'' if $Pr_> 70$.  The second test periodogram  < 2.8

\section{Some astrophysical implications}

\subsection {Masses}
\label{Sect:masses}

Masses of the components of spectroscopic binaries with one visible 
spectrum (SB1) are
encapsulated in the mass function
\begin{equation}
f(M_1,M_2) \equiv \frac{M_2^3 \sin^3 i}{(M_1 + M_2)^2} \equiv Q \sin^3 i,
\end{equation}
where $M_1$ and $M_2$ are the masses of the primary and secondary 
components, respectively.
The knowledge of the inclination as given in Table~\ref{Tab:NewOrb} 
gives directly access to the
generalized mass ratio $Q$ listed in Table~\ref{Tab:Mass}.  To go one 
step further and have access to
the masses themselves, supplementary  information must be injected in 
the process. For
main-sequence stars, this may come from the mass -- luminosity relationship.
The mass of the main-sequence primary component is estimated directly 
from its Hipparcos
$B-V$ color index, converted into an absolute magnitude $M_V$ using 
Table~15.7
of \citet{Cox}, and then into masses using Table~19.18 of \citet{Cox}.
The corresponding masses are
listed in Table~\ref{Tab:Mass}. The major uncertainty  on $M_2$ comes 
from the uncertainty on $M_1$
rather than from  $i$. To fix the ideas, an uncertainty of 0.1~mag on 
$B-V$ translates into an
uncertainty of  0.2 (or 0.1, 0.05) \Msun\ on $M_1$, and of 0.045 (0.032, 
0.027) \Msun\ on $M_2$ for
$0 \le M_V < 4$ (or $4 \le M_V < 6$, $6 \le M_V$, respectively). The 
position of stars from
Table~\ref{Tab:NewOrb} on the main sequence has been checked from the 
Hertzsprung-Russell diagram
drawn from the Hipparcos data. In particular, it has been checked that
the $B-V$ color is not the composite of the two components (in which
case, the above procedure to derive $M_1$ may not be applied). Only
HIP~47461 (= HD~83270) belongs to that category \citep[as confirmed by
   ][]{Ginestet-1991}, so that neither masses
are given in Table~\ref{Tab:Mass}.

Individual systems of interest are discussed in
Sect.~\ref{Sect:individualM}.

\begin{figure}[htb]
\resizebox{\hsize}{!}{\includegraphics{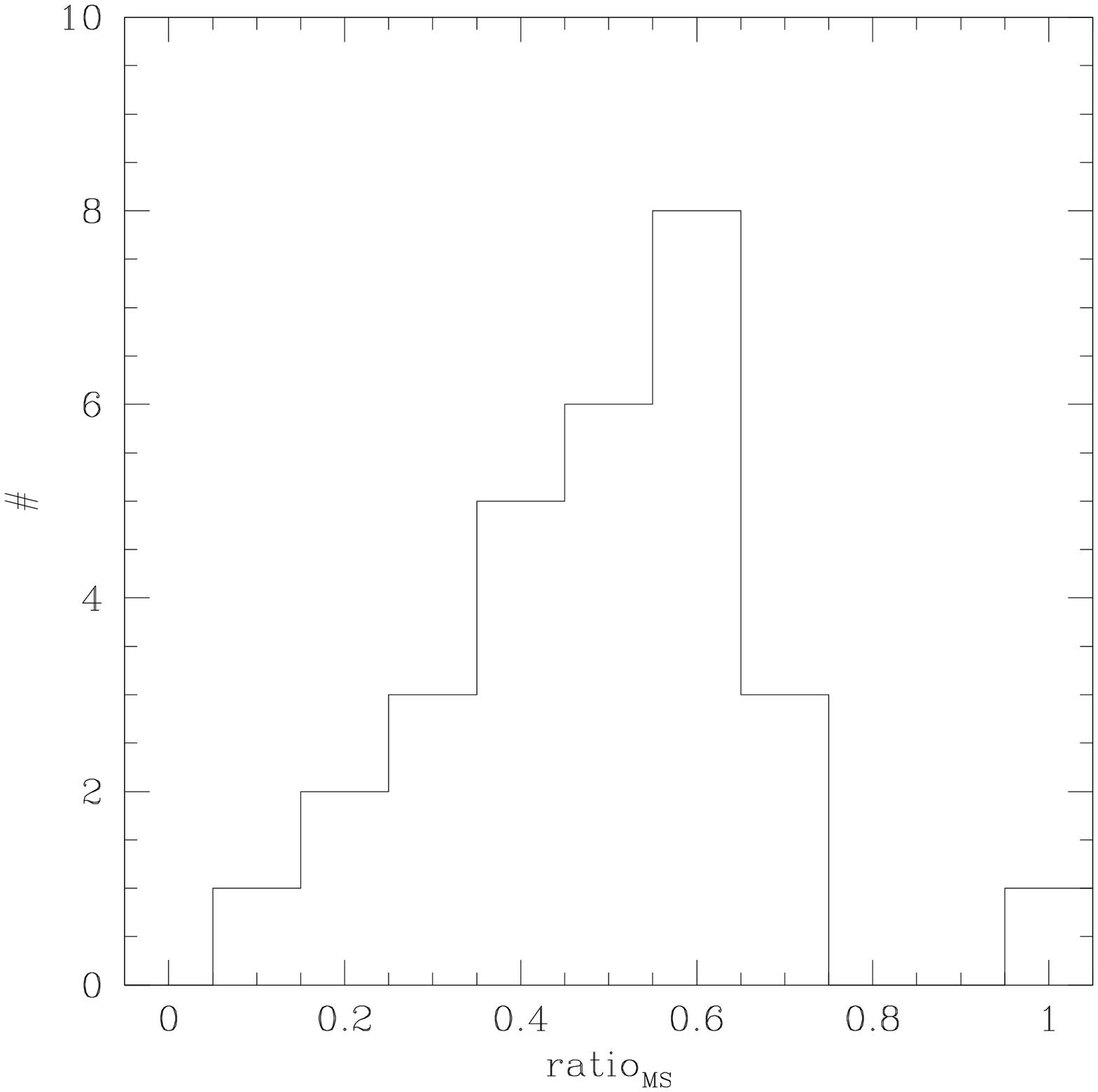}}
\resizebox{\hsize}{!}{\includegraphics{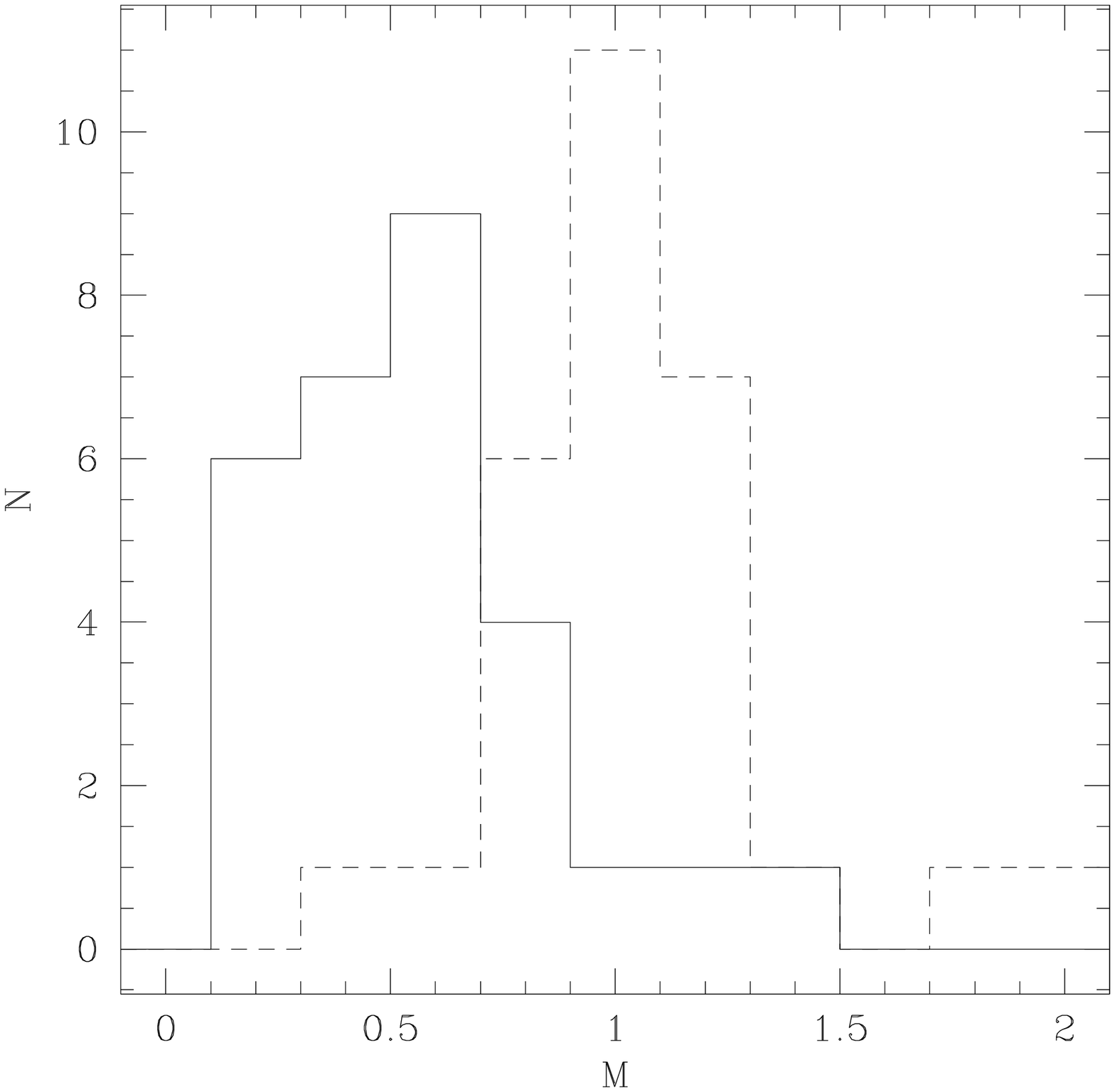}}
\caption{\label{Fig:masses} Upper panel: Distribution of the mass ratio 
($M_2/M_1$) for systems from Table~\ref{Tab:NewOrb} with a main-sequence
primary star. Lower panel: Distributions of $M_1$ (dashed  line)
and
$M_2$ (solid line).}
\end{figure}

The distributions of $M_1$, $M_2$ and $q = M_2/M_1$ for the 29 systems 
with main sequence primaries
are displayed in Fig.~\ref{Fig:masses}. The $q$ distribution appears to 
be strongly peaked around $q = 0.6$, but this feature very likely 
results from the combination of two opposite selection biases. Our 
sample is biased against systems with $q \sim 1$ (since these systems 
would generally be SB2 systems with components of almost equal 
brightness, whose astrometric motion is difficult to detect; see the 
discussion of Sect.~\ref{sect:res}) and against systems with low-mass 
companions (which induce radial-velocity variations of small amplitude, 
difficult to detect, and thus not present in \SB9).

The $M_1$ and $M_2$ distributions also clearly reflect the bias against 
$q = 1$ since the distributions exhibit adjacent peaks. Although one 
would be tempted to attribute the $M_2 = 0.6$ \Msun\ peak to a 
population of white dwarf (WD) companions, it is more likely to result 
from the two selection biases described above.

In the absence of a mass -- luminosity relationship for giants, the mass 
of the companion cannot be derived reliably.
%In four cases discussed below 
%(HIP~677; HIP~50801; HIP~77678; HIP~105017), 
%the mass of the giant could
%nevertheless be  estimated from various other astrophysical
%considerations.

\subsubsection {Masses for some specific systems}
\label{Sect:individualM}

\noindent{\it HIP 677 = }$\alpha$~{\it And}
\medskip\\
As already discussed in Sect.~\ref{Sect:orbit}, HIP~677  is known to be 
a SB2 and
visual binary \citep{Ryabchikova-1999,
Pourbaix-2000}. Masses are thus already available in the literature, 
namely $M_1 = 3.6\pm0.2$~\Msun,  $M_2 = 1.78\pm0.08$~\Msun\ 
\citep{Ryabchikova-1999}
or $M_1 = 3.85\pm0.22$~\Msun,
$M_2 = 1.63\pm0.074$~\Msun\ \citep{Pourbaix-2000}.
% At some phases of its orbital cycle, HIP~677
%($\alpha$~And)  clearly appears as a SB2 system, yielding $K_1 = 
%27.74$~kms and $K_2 = 65.46$~kms
%\citep{Pourbaix-2000}, yielding
%$M_1 \sin^3i = 3.45$~\Msun\ and $M_2 \sin^3i = 1.46$~\Msun. The orbital 
%inclination provided in
%Table~\ref{Tab:NewOrb} is then used to derive the individual masses 
%listed in %Table~\ref{Tab:Mass}.
\medskip\\
\noindent{\it HIP 20935 = HD 28394}
\medskip\\
This F7V star is a member of the Hyades cluster. It has a mass ratio $q
= M_2/M_1$ of 0.98. However, it falls exactly along the main sequence
as defined by the other stars of our sample. There is thus no
indication that this star has composite colors, as it should if the
companion is a main sequence F star as well. A white dwarf  (WD) companion
of mass  1.1~\Msun\
is not without problems either.
\citet{Bohm-Vitense-1995} has searched the IUE {\it International Ultraviolet Explorer}
 archives for spectra of
F stars from the Hyades, in order to look for possible WD
companions. No excess UV flux is present at 142.5 nm for HIP~20935,
which implies that the WD must be cooler than about 10000~K. For a
1.1~\Msun\ WD, this implies a cooling time of more than 1~Gyr
\citep{Chabrier-2000}, incompatible with the Hyades age of $800\;10^6$~y.
The remaining possibility is that the companion is itself a binary with two
low-luminosity red dwarfs.
\medskip\\
\noindent {\it HIP 105969 = HD 204613}
\medskip\\
This star is known as a subgiant CH star \citep{Luck-Bond-1982} and the 
system should therefore host a WD companion \citep{McClure-1997}. 
Interestingly enough, the mass inferred for this WD companion is 
0.49~\Msun, just large enough for a 2~\Msun\ AGB progenitor to have gone 
through the thermally-pulsing asymptotic granch branch phase \citep[see 
Fig.~3.10 in][]{Groenewegen-2003} to synthesize heavy elements by the 
s-process of nucleosynthesis. Those heavy elements were subsequently 
dumped onto the companion (the current CH subgiant) through mass transfer.

\begin{table*}[bth]
\caption[]{\label{Tab:Mass} Masses and mass ratios for the 29 systems 
with main-sequence primaries
passing all consistency tests.}
\begin{tabular}{rcccccp{5cm}}
\hline
\noalign{\smallskip}
HIP & $M_1$ (\Msun)  & $M_2$(\Msun) &  $M_2/M_1$ & $Q =
\frac{M_2^3}{(M_1+M_2)^2}$ & Sp. Type & Rem. \\
\noalign{\smallskip}
\hline
\noalign{\smallskip}
%\noalign{Main sequence stars}
1349 & 0.98 & 0.55 & 0.56 & 0.0711 & G2 & \\
1955 & 1.13 & 0.48 & 0.42 & 0.0420 & GO & \\
7078 & 1.21 & 0.70 & 0.58 & 0.0953 & F6 & \\
8903 & 1.86 & 1.05 & 0.56 & 0.1358 & A5 & \\
11231 &1.01 & 0.68 & 0.67 &  0.111 & G2 & \\
\\
12062 & 0.95 & 0.44 & 0.47 & 0.0449 & G5 & \\
20935 & 1.13 & 1.11 & 0.98 & 0.2732 & F7 & not a composite spectrum
despite a mass ratio close to unity\\ 
24419 & 0.90 & 0.21 & 0.24 & 0.0079
& G8 &
\\ 34164 & 1.09 & 0.66 & 0.61 & 0.0954 & G0& \\
%34608 & 0.86 & 0.63 & 0.73 & 0.1115 & G9& sous-geante\\
39893 & 0.95 & 0.52 & 0.55 & 0.0649 &  G3& \\
\\
47461 & -    & -    & -    & 0.0863 & F2& composite spectrum\\
%59750 & 1.21 & 0.53 & 0.44 & 0.0484 & F5& \\
63406 & 0.82 & 0.23 & 0.28 & 0.0114 & G9& \\
72848 & 0.79 & 0.45 & 0.56 & 0.0581 & K2& \\
73440 & 1.03 & 0.15 & 0.14 & 0.0023 & G0& \\
75379 & 1.26 & 0.68 & 0.54 & 0.0842 & F5 & \\
\\
%77409 & 0.69 & 0.18 & 0.27 & 0.0082 & K5 & \\
79101  & 3.47 &  1.31 &  0.38 & 0.0976 & B9 & \\
80346 & 0.50 & 0.13 & 0.26 & 0.0054 &M3 & \\
80686 & 1.05 & 0.37 &  0.36 &  0.0259 & G0 & \\
82860 & 1.18 & 0.52 & 0.44 & 0.0482 & F6 & \\
86400 & 0.72 & 0.39 & 0.54 & 0.0475 & K3& \\
87895 & 0.99 & 0.68 & 0.69 & 0.1129 & G2& \\
\\
89937 & 1.18 & 0.77 & 0.65 & 0.1195 & F7Vvar& \\
95028 & 1.40 & 0.50 &  0.36 & 0.0353 & F5 & \\
95575 & 0.78 & 0.38 & 0.49 & 0.0405 & K3& \\
99965 & 0.88 & 0.56 & 0.63 & 0.0840 & G5& \\
%105860 & ??\\
105969 &  1.01 & 0.49 &  0.49 &  0.0528 & Dwarf Ba/Subgiant CH & \\
109176 & 1.25 & 0.80 & 0.64 & 0.1233 & F5 & \\
\\
111170 & 1.08 & 0.70 & 0.65 & 0.1083 & F7 & \\
113718 & 0.76 & 0.18 & 0.24 & 0.0067 & K4 & \\
%\medskip\\
%\noalign{Giant}\\
%\medskip\\
%677  & $3.85\pm0.22$ & $1.63\pm0.074$ & 0.42& 0.140 &B8IV& individual 
%masses derived from SB2 data (see text)\\
%50801& 4.0& 0.82& 0.20& 0.024 &M0III & the "hybrid" nature of the giant 
%points toward $M_1
%\sim 4$~\Msun \citep{Reimers-1996}\\
%77678& 3.8& 3.00& 0.79& 0.588 & K2III + A0V & $M_V$(2) = 0.6 (companion 
%assumed to be a main sequence
%star)\\
%      & 2.6& 2.50& 0.96& 0.588 & K2III + A3V & $M_V$(2) = 1.5 (preferred 
%solution; see
%text)\\
%      & 1.7& 2.00& 1.18& 0.588 & K2III + A5V & $M_V$(2) = 1.9 \\
%      & 1.5& 1.89& 1.26& 0.588 & K2III + A7V & $M_V$(2) = 2.2 \\
%105017& 2.5& 1.8& 0.72& 0.310 &K0III + A7V& masses chosen to be 
%consistent with companion's spectral
%type\\
\noalign{\smallskip}
\hline
\end{tabular}
\end{table*}

\subsection {\elogP\ diagram}
\label{Sect:elogP}

With the availability of extensive sets of orbital elements for binaries 
of various kinds ({\it
e.g.}, Duquennoy \& Mayor 1991 for G dwarfs, Matthieu 1992 
\nocite{Matthieu-1992} for pre-main sequence
binaries, Mermilliod 1996 for open-cluster giants, Carney et al. 2001 
for blue-straggler,
low-metallicity stars, Latham et al. 2002 for halo stars),   it has 
become evident that
long-period ($P > 100$~d), low-eccentricity ($e < 0.1$) systems are 
never found among unevolved
({\it i.e.,} pre-mass-transfer) systems. This indicates that binary 
systems always form in eccentric
orbits, and the shortest-period systems are subsequently circularized by 
tidal effects. On the
contrary,  binary systems  which can be ascribed post-mass-transfer 
status because they exhibit
signatures of chemical pollution due to mass transfer (like barium 
stars, some subgiant CH stars,
S stars without technetium lines...) are often found in the avoidance 
region ($P > 100$~d,
$e < 0.1$) of the \elogP\ diagram. Mass transfer indeed severely 
modifies their orbital elements,
which often end up in this region \citep{Jorissen-03, Jorissen-05}.

\begin{figure}[htb]
\resizebox{\hsize}{!}{\includegraphics{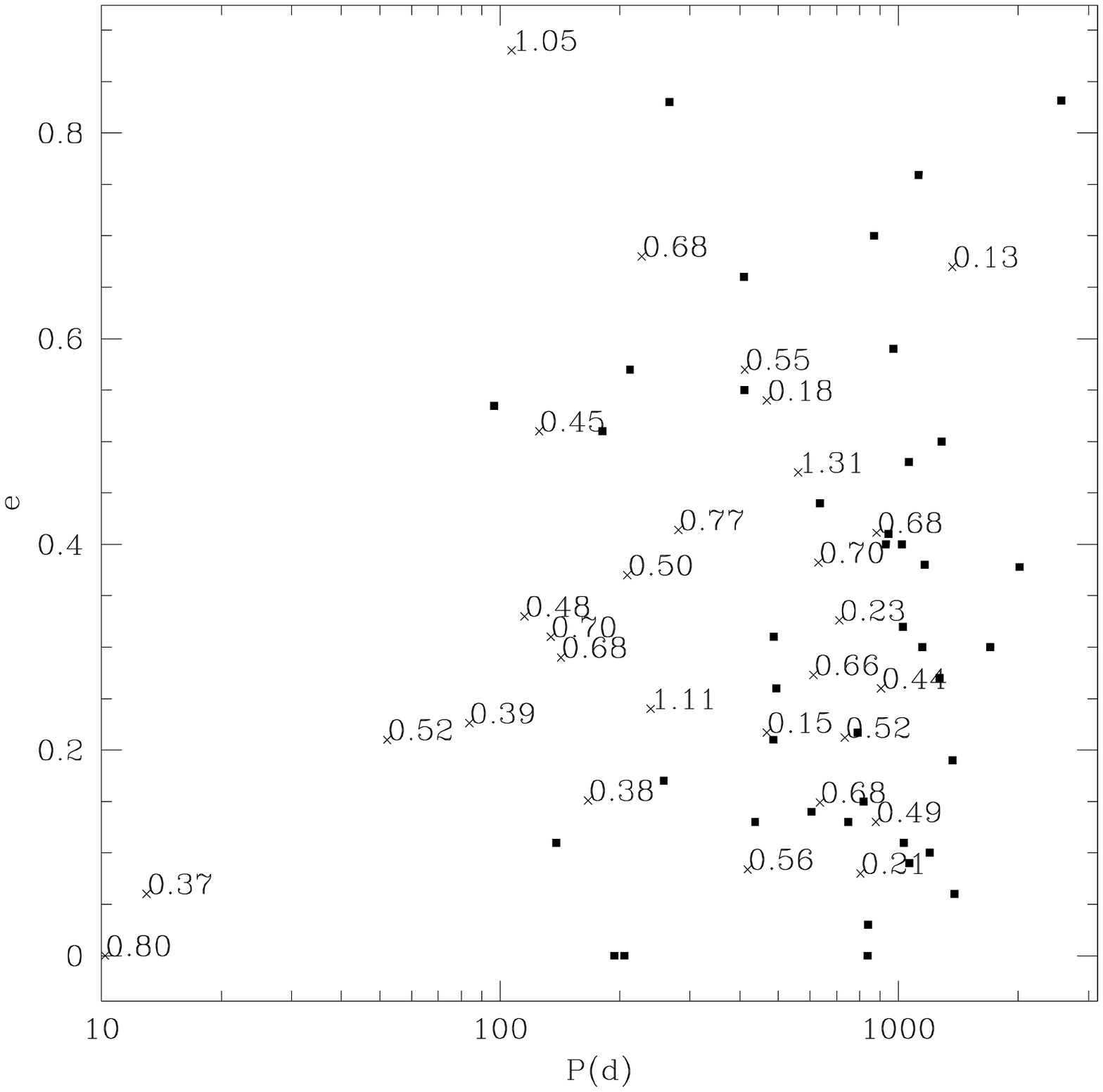}}
\caption[]{\label{Fig:elogPMass}
The \protect\elogP\ diagram for the 70 systems with reliable astrometric 
orbital elements.
Systems with giant primaries are represented by black squares, and 
main-sequence primaries with
crosses. The point labels refer to the companion mass.
}
\end{figure}

The companion masses derived in Sect.~\ref{Sect:masses} offer the 
opportunity to check whether
  systems falling in the avoidance region of the \elogP\ diagram could be
post-mass transfer systems (most probably then with a WD companion).
In total, 8 systems fall in this region, as displayed on 
Fig.~\ref{Fig:elogPMass}: HIP~6867 (= HD~9053 = $\gamma$~Phe; M0~III),
HIP~8922 (= HD~11613 = HR~551; K2), HIP~10514 (= HD~13738; 
K3.5~III), HIP~24419 (= HD~34101; G8~V), HIP~32768 (= HD~50310 = HR 2553; K1~III), 
%HIP~59750 (F5~V), 
HIP~99965 (= HD~193216; G5~V), HIP~101093 (= HD~195725; A7~III) and HIP~101847 (= HD~196574; G8~III).

%Among these, HIP~59750 (= HD~106516, F5V) is the only system to offer 
%conclusive evidence for a WD companion. It is a soft X-ray source [$L_X 
%= 1.4
%\pm0.7\; 10^{28}$ erg/s with a ROSAT hardness ratio of
%$ (H-S) / (H+S) = -0.54$, where $H$ and $S$ are the countrates in the 
%hard (0.5 --
%2.0 keV) and soft (0.1 -- 0.28 keV) bandpasses, respectively] 
%\citep{Ottmann-1997}, which  could be
%indicative of a WD, consistent with its inferred mass of 0.53~\Msun. 
%This case is especially
%interesting as there are two conflicting orbital periods in the 
%literature: a short one
%\citep[23.1~d;][]{Abt-Willmarth-1987} and a long one
%\citep[843.9~d;][]{Carney-2001}. The fact that the astrometric data 
%yield a consistent solution with
%the long-period orbit certainly gives strong credit to it. In their 
%discussion of the X-ray
%properties of this star, \citet{Ottmann-1997} adopted instead the short 
%period; in that case, the
%coronal origin of the X-rays was favoured, as a result of the  star's 
%suspectedly fast rotation due
%to orbital synchronization. With the long-period orbit, this explanation 
%is no more possible,
%however. The softness of the X-rays, the very value of
%$L_X$, the blue straggler nature of HIP~59750
%\citep{Carney-2001}, and finally the mass of the companion (0.53~\Msun; 
%Table~\ref{Tab:Mass}) all point instead towards a WD companion.

None of these 'avoidance-region' systems offer conclusive evidence for hosting 
a WD companion, but at least do not contradict it either.

%the hypothesis that the `avoidance region' of the 
%eccentricity -- period diagram is populated by post-mass-transfer system.
HIP~6867
has a circular orbit and a rather short orbital period (193.8~d) given 
its late spectral type. The orbit is therefore likely to have been 
circularized by tidal effects rather than by mass transfer 
\citep{Jorissen-2004}. In this specific case, there is therefore no need for the 
companion to be a WD. 

HIP~24419 has too small a companion mass (0.21~\Msun) 
to host even a He WD.  This system could nevertheless have gone through 
a so-called `case B' mass transfer (occurring when the primary was on 
the first giant branch).

For HIP~8922, HIP~10514, HIP~32768, HIP~99965,
HIP~101093 and HIP~101847, we could not find in the literature any 
information that could help us in assessing the nature of their 
companion. In the case of HIP~99965 though, the companion's mass of 
0.56~\Msun\ would certainly not dismiss it of being a WD.
%HIP~50801 is a well-documented hybrid star
%5\citep[characterized by the simultaneous presence of cool winds and hot 
%transition region and corona; ] []{Reimers-1992, Reimers-1996}, with an 
%X-ray luminosity of $3.5\; 10^{28}$~erg/s possibly caused by a corona 
%with a temperature of $3\;10^6$~K. For this system, there is thus no 
%strong evidence in favour of a WD companion, although nothing excludes 
%it \citep{Reimers-1992}.

One should mention as well that HIP~10514 and HIP~101847 are listed in the {\it Perkins 
catalog of revised MK types for the cooler stars} \citep{Keenan-1989} 
without any mention whatsoever of spectral peculiarities. They are 
therefore definitely not barium stars, despite falling in the 
'avoidance region' of the eccentricity -- period diagram generally 
populated by barium stars. If we are to maintain that the 'avoidance 
region' can only be populated by post-mass-transfer objects -- thus 
implying that the companion to HIP~10514 and all the stars discussed in 
the present section {\it must} be WDs  -- then we must accept at the 
same time that systems following the same binary evolution channel as 
that of barium stars do not necessarily end up as barium stars! Or in 
other words, binarity would not be a sufficient condition for the barium 
syndrome to develop \citep[these systems would thus add to the 
non-barium binary systems listed in ][]{Jorissen-Boffin-1992}.
%
%------------------------------------------------------------------------------
\section{Conclusions}
%------------------------------------------------------------------------------
%

The major result of this paper is that the detectability of an 
astrometric binary using the IAD is
mainly a function of the orbital period (at least when the parallax 
exceeds 5~mas, {\it i.e.,} about 5
times the standard error on the parallax): detection rates are close to 
100\% in the period range
50 -- 1000~d (corresponding to the mission duration) for systems {\it 
not} involving
components with almost equal brightnesses ({\it i.e.,} SB2 systems or 
systems with composite spectra).
These are more difficult to detect, because the photocenter motion is 
then much smaller than the actual
component's motion.

A consistency test between Thiele-Innes and Campbell solutions has been 
designed that allowed us to
(i) identify wrong spectroscopic solutions, and (ii) retain  70 systems 
with accurate
orbital inclinations (among those, 29 involve main sequence primaries 
and 41 giant primaries). Among
those 70 retained solutions, 20 are new astrometric binaries, not listed 
in the DMSA/O.

This number of 70 systems passing all quality checks seems small with 
respect to the 122 DMSA/O
systems with an \SB9\ entry. A detailed check reveals, however, that 
many systems present in the
DMSA/O either have inaccurate astrometric orbits  that would not fulfill 
our statistical tests, or have
inaccurate spectroscopic orbital elements that make the astrometric 
solution unreliable anyway, or
have only $a_0$ derived from the IAD, all other elements being taken 
from spectroscopic and
interferometric/visual orbital elements.

Masses $M_2$ for the companions in the 29 systems hosting a 
main-sequence primary star have been
derived, using the mass-luminosity relation to estimate $M_1$.  This was 
not possible for systems
hosting giant primaries. 
%One system (HIP~77678, K2~III) has, however, 
%well-constrained masses
%if one imposes $q = M_2/M_1 < 1$: $M_1 = 2.8\pm0.2$~\Msun\ and $M_2 = 
%2.6\pm0.1$~\Msun, yielding $q
%\sim 0.93$.  A purely dynamical mass has been obtained for the WD 
%companion in the Sirius-like system
%HIP~105860 (=IK~Peg = A8m + hot WD), amounting to  1.3~\Msun\ and 
%contrasting with the value
%$0.985\pm0.03$~\Msun\  obtained from spectral fits to X-ray and EUV 
%data. This would make it one
%among the more massive WDs known.

The possibility that the region $e < 0.1$, $P > 100$~d of the \elogP\ 
diagram is exclusively
populated by post-mass transfer systems has been examined, but could not 
be firmly demonstrated. 
%There
%are indeed strong indications that at least some among the systems 
%populating that region of the
%\elogP\ diagram have WD companions (HIP 59750), but the evidence is far 
%from being as strong for
%the other systems.

\begin{acknowledgements}
AJ and DP are Research Associates, FNRS (Belgium).  This research was 
supported in part by the
ESA/PRO\-DEX Research Grants 90078 and 15152/01/NL/SFe. We thank M. 
Hallin and
A. Albert for discussions.
We would like
to thank the referee of this paper,
Prof. L. Lindegren, for his very valuable comments and suggestions.
\end{acknowledgements}

\bibliographystyle{aa}
\bibliography{3003arti,3003book}

\end{document}